\documentclass[dvips]{article}
\usepackage{supertabular,lscape,epsfig}
\usepackage{amssymb}
\usepackage{amsmath}

\DeclareSymbolFont{ppa}{OT1}{ppl}{m}{it}
\DeclareMathSymbol{\vv}{\mathalpha}{ppa}{'166}

\begin{document}

\newcommand{\dd}{\,{\rm d}}
\newcommand{\ie}{{\it i.e.},\,}
\newcommand{\etal}{{\it et al.\ }}
\newcommand{\eg}{{\it e.g.},\,}
\newcommand{\cf}{{\it cf.\ }}
\newcommand{\vs}{{\it vs.\ }}
\newcommand{\zdot}{\makebox[0pt][l]{.}}
\newcommand{\up}[1]{\ifmmode^{\rm #1}\else$^{\rm #1}$\fi}
\newcommand{\dn}[1]{\ifmmode_{\rm #1}\else$_{\rm #1}$\fi}
\newcommand{\upd}{\up{d}}
\newcommand{\uph}{\up{h}}
\newcommand{\upm}{\up{m}}
\newcommand{\ups}{\up{s}}
\newcommand{\arcd}{\ifmmode^{\circ}\else$^{\circ}$\fi}
\newcommand{\arcm}{\ifmmode{'}\else$'$\fi}
\newcommand{\arcs}{\ifmmode{''}\else$''$\fi}
\newcommand{\MS}{{\rm M}\ifmmode_{\odot}\else$_{\odot}$\fi}
\newcommand{\RS}{{\rm R}\ifmmode_{\odot}\else$_{\odot}$\fi}
\newcommand{\LS}{{\rm L}\ifmmode_{\odot}\else$_{\odot}$\fi}

\newcommand{\Abstract}[2]{{\footnotesize\begin{center}ABSTRACT\end{center}
\vspace{1mm}\par#1\par
\noindent
{~}{\it #2}}}

\newcommand{\TabCap}[2]{\begin{center}\parbox[t]{#1}{\begin{center}
  \small {\spaceskip 2pt plus 1pt minus 1pt T a b l e}
  \refstepcounter{table}\thetable \\[2mm]
  \footnotesize #2 \end{center}}\end{center}}

\newcommand{\TableSep}[2]{\begin{table}[p]\vspace{#1}
\TabCap{#2}\end{table}}

\newcommand{\FigCap}[1]{\footnotesize\par\noindent Fig.\  %
  \refstepcounter{figure}\thefigure. #1\par}

\newcommand{\TableFont}{\footnotesize}
\newcommand{\TableFontIt}{\ttit}
\newcommand{\SetTableFont}[1]{\renewcommand{\TableFont}{#1}}

\newcommand{\MakeTable}[4]{\begin{table}[htb]\TabCap{#2}{#3}
  \begin{center} \TableFont \begin{tabular}{#1} #4 
  \end{tabular}\end{center}\end{table}}

\newcommand{\MakeTableSep}[4]{\begin{table}[p]\TabCap{#2}{#3}
  \begin{center} \TableFont \begin{tabular}{#1} #4 
  \end{tabular}\end{center}\end{table}}

\newenvironment{references}%
{
\footnotesize \frenchspacing
\renewcommand{\thesection}{}
\renewcommand{\in}{{\rm in }}
\renewcommand{\AA}{Astron.\ Astrophys.}
\newcommand{\AAS}{Astron.~Astrophys.~Suppl.~Ser.}
\newcommand{\ApJ}{Astrophys.\ J.}
\newcommand{\ApJS}{Astrophys.\ J.~Suppl.~Ser.}
\newcommand{\ApJL}{Astrophys.\ J.~Letters}
\newcommand{\AJ}{Astron.\ J.}
\newcommand{\IBVS}{IBVS}
\newcommand{\PASP}{P.A.S.P.}
\newcommand{\Acta}{Acta Astron.}
\newcommand{\MNRAS}{MNRAS}
\renewcommand{\and}{{\rm and }}
\section{{\rm REFERENCES}}
\sloppy \hyphenpenalty10000
\begin{list}{}{\leftmargin1cm\listparindent-1cm
\itemindent\listparindent\parsep0pt\itemsep0pt}}%
{\end{list}\vspace{2mm}}

\def\TYLDA{~}
\newlength{\DW}
\settowidth{\DW}{0}
\newcommand{\dw}{\hspace{\DW}}

\newcommand{\refitem}[5]{\item[]{#1} #2%
\def\REFARG{#3}\ifx\REFARG\TYLDA\else, {\it#3}\fi
\def\REFARG{#4}\ifx\REFARG\TYLDA\else, {\bf#4}\fi
\def\REFARG{#5}\ifx\REFARG\TYLDA\else, {#5}\fi.}

\newcommand{\Section}[1]{\section{#1}}
\newcommand{\Subsection}[1]{\subsection{#1}}
\newcommand{\Acknow}[1]{\par\vspace{5mm}{\bf Acknowledgements.} #1}
\pagestyle{myheadings}

\newfont{\bb}{ptmbi8t at 12pt}
\newcommand{\xrule}{\rule{0pt}{2.5ex}}
\newcommand{\xxrule}{\rule[-1.8ex]{0pt}{4.5ex}}
\def\thefootnote{\fnsymbol{footnote}}
\begin{center}
{\Large\bf Multimode Resonant Coupling in Pulsating Stars.\\
\vskip3pt}
\vskip1cm
{Rafa\l~~M.~~N~o~w~a~k~o~w~s~k~i}
\vskip3mm
Nicolaus Copernicus Astronomical Center, ul.~Bartycka~18, 
00-716~Warsaw, Poland\\
e-mail: rafaln@camk.edu.pl

\end{center}

\Abstract{We consider evolution of an unstable acoustic mode
interacting with an ensemble of stable g-modes. We show that the static
multimode solution does not exist. We then find the condition for the
stability of the statistical equilibrium.

Performing numerical integration of amplitude equations for a simplified
system we find that the acoustic mode amplitude exhibits a large irregular
variability on the timescale given by the inverse of the growth rate.

The g-mode pairs are excited in significantly wider range of detuning
parameters than it is implied by the parametric instability criterion applied
to the average amplitude. However, the number of interacting g-mode pairs
is reduced because the pairs differing in the
detuning parameter by less than their damping rates are synchronized and
effectively act as a single pair.

We apply the multimode resonant coupling theory to a realistic stellar
model. We choose a seismic model of the $\delta$-Scuti star XX~Pyxidis.
Although for some $l=2$ modes we find amplitudes of the order of a few
milimagnitudes, the typical amplitudes of low-degree modes are much higher.
Taking into account the rotational splitting
results in decrease of amplitudes by a factor of few
which is not enough to obtain consistency with observations. We conclude that
in this star and likely in all evolved $\delta$-Scuti stars,
the resonant mode
coupling cannot be the dominant amplitude limiting effect. The nonresonant
saturation of the driving effect must play the role.
}
{Stars: oscillations -- Stars: variables: $\delta$~Scuti}

\vspace*{12pt}
\Section{Introduction}

In our understanding of stellar multiperiodic variability, 
which is the typical form of pulsation in main sequence stars and 
evolved dwarfs, we rely mostly on  the linear  approximation. 
This suffices if our aim is calculation of oscillation mode
frequencies. Taking into account 
nonadiabatic effects allows to determine which modes are linearly unstable.
This instability usually happens in some ranges of spherical harmonic degrees
$l$ and eigenfrequencies. There is fairly good agreement between theoretical
frequency ranges of unstable modes and frequencies present in real stars.
However, the big unsolved problem is why most of the theoretically unstable 
modes are not seen in observations. This problem is related to the more
general problem
of amplitude limitation mechanism, which needs nonlinear approach.

The problem with mode amplitudes concerns in particular $\delta$~Scuti stars.
Despite the same excitation mechanism as in the giant classical pulsators, 
namely the $\kappa$ mechanism, their pulsational behavior is very different
and more diverse.

First of all, it was noticed already in seventies (Breger 1979) that only 
about 30\% of stars falling into lower instability strip show detectable
variability. This number is somewhat higher today but still less then
one half of the stars in the $\delta$~Scuti domain are detected as pulsating
(Breger 2000a). The large number of variables with amplitudes close to the 
detection limit suggest that many (perhaps all) of the stars not detected as
variable also pulsate but with very small amplitudes.

There is an 
important subgroup of high-amplitude $\delta$~Scuti stars (HADS). Their
pulsational properties resemble those of Cepheid-type variables, \ie only one
or two lowest radial modes are present with photometric amplitudes of the order
of $0.5$ mag. These stars are slowly rotating ($v\sin i\lesssim30$ km/s) 
and the oscillation amplitudes are constant or almost constant (see, \eg
Rodriguez 1999). They are more evolved than the rest of the class. However,
the majority of $\delta$~Scuti stars are low-amplitude pulsators with many
independent modes most of which are nonradial. These typical $\delta$~Scuti
stars can rotate quite fast, with $v\sin i$ up to $250$ km/s, but some of them
are also slow rotators (Breger 2000a). The pulsation amplitudes of the 
dominant modes are 
below $0.1$ mag, and those of the remaining modes are usually less than 10 
mmag.

Many low-amplitude $\delta$~Scuti stars show long-time amplitude modulations.
Breger (2000b) analyzed over 30 years of observational data of evolved star
4CVn. The amplitudes of most modes in this star change significantly in the 
time-scale of ten years or longer. One mode showed rapid decrease of 
amplitude from 15 mmag in 1974 to 4 mmag in 1976 and 1 mmag in 1977 and later
started growing. There was also a phase jump between 1976 and 1977. The amount
of amplitude variability in other modes was of the order of 40\% during a
decade.

A different example is the star XX~Pyx that was observed by Handler \etal 
(2000).
The observations covered much shorter time, only 125 nights, but due to 
continuous observation at eight observatories spread around the world and a
sophisticated method of nonlinear frequency analysis it was possible to
detect amplitude and phase modulations of many modes with the time-scales 
from 20 to 70 days.

The possible explanation of the apparent difference between pulsational 
behavior of giant-type pulsators and typical $\delta$~Scuti stars involves
the resonant mode coupling. The interaction between one higher 
frequency wave and two lower
frequency waves is well known in plasma physics (see, \eg Davidson 1972). 
A detailed analysis of general solutions is given by Wersinger \etal 
(1980). In the case of stellar pulsations it was first studied by Vandakurov
(1965, 1979). Dziembowski (1982) developed a mathematical formalism suitable
to study this kind of interaction in $\delta$~Scuti stars. Application of this
formalism to a  model of ZAMS $\delta$~Scuti star (Dziembowski and
Kr{\'o}likowska 1985)
suggested that the resonant mode coupling indeed was
an effective mechanism of amplitude limitation.
Dziembowski \etal (1988) studied the effects of rotation and showed that the
amplitudes could be even smaller than in the nonrotating case.
However, all these studies concerned a stationary situation. In fact, often
happens that the stationary solutions are unstable and the dynamical solutions
need to be investigated. It was done by Moskalik (1985) in the case of 
three-mode interaction. The realistic many-mode case was studied only in the 
context of tidal-capture binaries (Kumar and Goodman, 1996). In that study,
the $l=2$ mode had been previously excited due to tidal interaction and
the nonlinear couplings caused damping of that mode. It is a different
situation than in the case of unstable acoustic modes in $\delta$-Scuti stars,
and we cannot strictly apply that study to our case. 

\Section{Parametric Resonance}

\Subsection{The Case of a Single G-Mode Pair}

\subsubsection{Amplitude Equations}

The amplitude equations describing resonant interaction between an acoustic 
mode with frequency $\omega_0$ and a pair of gravity modes whose frequencies, 
$\omega_1, \omega_2$, satisfy the condition 
$|\omega_0-\omega_1-\omega_2|\ll\omega_{0,1,2}$ are derived, \eg by
Dziembowski (1982):
\begin{eqnarray}
\frac{{\rm d} A_0}{{\rm d} \tau}&=&\gamma_0 A_0+{\rm i}\frac{\epsilon H}
{2\sigma_0 I_0}A_1 A_2 {\rm e}^{-{\rm i}\Delta\sigma\tau},\\
\frac{{\rm d} A_{1,2}}{{\rm d} \tau}&=&\gamma_{1,2} A_{1,2}+{\rm i}\frac{H}
{2\sigma_{1,2} I_{1,2}}A_0 A_{2,1}^* {\rm e}^{{\rm i}\Delta\sigma\tau},
\end{eqnarray}
where $A$ are dimensionless complex amplitudes of the interacting modes at
the surface, $\gamma$ denote 
growth rates, $I$ are moments of inertia, $H$ is the coupling coefficient, and
the dimensionless time, $\tau$, and frequencies, $\sigma$, are related to
the real ones, $t,\omega$ by the relation 
$\tau/t=\omega/\sigma=\sqrt{4\pi G<\rho>}$. The frequency mismatch is given by
$\Delta\sigma\equiv\sigma_0-\sigma_1-\sigma_2$, and $\epsilon$ distinguishes 
the case of two-mode resonance, where the subscripts $1,2$ denote the same 
mode, and the case of three-mode resonance, where they denote different modes.
In the former case $\epsilon=1/2$ and in the latter case $\epsilon=1$.
The acoustic mode growth rate is positive and the g-mode growth rates are
negative.

Dziembowski (1982) made an explicit use of adiabatic assumption with 
artificially added growth rates. This yields real
coupling coefficient. However, following a derivation procedure 
of Van~Hoolst (1994b), we find the nonadiabatic amplitude equations in very 
similar form with the complex $\mathcal H$
\begin{eqnarray}
\frac{{\rm d} A_0}{{\rm d} \tau}&=&\gamma_0 A_0+{\rm i}
\frac{\epsilon\mathcal H}{2\sigma_0 I_0}
A_1 A_2 {\rm e}^{-{\rm i}\Delta\sigma\tau},\nonumber\\
\frac{{\rm d} A_{1,2}}{{\rm d} \tau}&=&\gamma_{1,2} A_{1,2}+{\rm i}
\frac{\mathcal H^*}{2\sigma_{1,2} I_{1,2}}A_0 A_{2,1}^* 
{\rm e}^{{\rm i}\Delta\sigma\tau}.\nonumber
\end{eqnarray}
If we put $\mathcal{H}=H{\rm e}^{{\rm i}\alpha}$ and shift complex phases of
g-mode amplitudes by $-{\rm i}\alpha/2$ the above amplitude equations are
reduced to Eqs.~(1),(2). Therefore, without loss of generality, we will use 
them with real and positive $H$. Furthermore, we rescale the coupling 
coefficient
\begin{equation}
C\equiv\frac{H}{\sqrt{\sigma_1\sigma_2 I_1 I_2}}
\end{equation}
and the g-mode amplitudes
\begin{equation}
A'_{1,2}\equiv A_{1,2}\sqrt{\frac{\epsilon\sigma_{1,2}I_{1,2}}
{\sigma_0 I_0}}
\end{equation}
in order to obtain simpler form of amplitude equations
\begin{eqnarray}
\frac{{\rm d} A_0}{{\rm d} \tau}&=&\gamma_0 A_0+{\rm i}\frac{C}{2}
A'_1 A'_2 {\rm e}^{-{\rm i}\Delta\sigma\tau},\\
\frac{{\rm d} A'_{1,2}}{{\rm d} \tau}&=&\gamma_{1,2} A'_{1,2}+{\rm i}\frac{C}
{2}A_0 A'_{2,1}{}^* {\rm e}^{{\rm i}\Delta\sigma\tau}.
\end{eqnarray}

There are various long timescales which play roles in the evolution of our
system. These are the detuning scale, $|\Delta\sigma|^{-1}$, the nonadiabatic
scales, $|\gamma|^{-1}$ and the coupling scale $|CA_0|^{-1}$.

If the amplitudes are sufficiently small then only linear terms play role and
the amplitude of the acoustic mode grows exponentially. Eventually it reaches
the point where the nonlinear terms in Eqs.~(6) become important.
Assuming sufficiently slow growth of the acoustic mode and using the procedure
of Dziembowski (1982) we obtain a criterion for the 
parametric excitation of the pair of linearly damped g-modes
\begin{equation}
|A_0|>B_{cr}\equiv\sqrt{\frac{4\gamma_1\gamma_2}{C^2}\left[1+
\left(\frac{\Delta\sigma}{\gamma_1+\gamma_2}\right)^2\right]}.
\end{equation}
When the acoustic mode amplitude exceeds the critical value the g-mode 
amplitudes start to grow and finally the nonlinear term in Eq.~(5) can stop
the exponential growth of the acoustic mode

In further applications it will be more suitable to use real form of amplitude 
equations. By substituting
\begin{equation}
A_0=B_0 {\rm e}^{{\rm i}\phi_0},\quad 
A'_{1,2}=B_{1,2}{\rm e}^{{\rm i}\phi_{1,2}}
\end{equation}
where $B_j$ are the real and positive amplitudes, and $\phi_j$ are the real 
phases of the considered modes, we obtain
\begin{eqnarray}
\frac{{\rm d} B_0}{{\rm d} \tau}&=&\gamma_0 B_0+\frac{C}{2}B_1 B_2 \sin\Phi,\\
\frac{{\rm d} B_{1,2}}{{\rm d} \tau}&=&\gamma_{1,2} B_{1,2}-\frac{C}{2}
B_0 B_{2,1} \sin\Phi,\\
\frac{{\rm d} \Phi}{{\rm d} \tau}&=&\Delta\sigma+\frac{C}{2}\left[\frac{
B_1B_2}{B_0}-B_0\left(\frac{B_2}{B_1}+\frac{B_1}{B_2}\right)\right]\cos\Phi,
\end{eqnarray}
where $\Phi\equiv\Delta\sigma\tau+\phi_0-\phi_1-\phi_2$.

\subsubsection{Equilibrium Solutions}

Here, we repeat the analysis of Dziembowski~(1982), who derived the formulae
for the equilibrium amplitudes and the criteria for the stability of the
equilibrium solution.

If we set l.h.s. of Eqs.~(9)--(11) equal to zero we find equilibrium solutions
\begin{eqnarray}
B_0^s&=&\sqrt{\frac{4\gamma_1\gamma_2}{C^2}\left(1+q^2\right)},\\
B_{1,2}^s&=&\sqrt{\frac{-4\gamma_0\gamma_{2,1}}{C^2}\left(1+q^2\right)},\\
\cot\Phi^s&=&q,
\end{eqnarray}
where $q=\Delta\sigma/\gamma$, and $\gamma=(\gamma_0+\gamma_1+\gamma_2)$.

The stationary solutions describe possible physical state only if they are 
stable to small perturbations. Linearizing Eqs.~(9)--(11) around equilibrium
solution (Eqs.~12--14) we have
\begin{equation}
\frac{\rm d}{{\rm d}\tau}\left[\begin{array}{c}\delta B_0/B_0\\
\delta B_1/B_1\\ \delta B_2/B_2\\ \delta\Phi\end{array}\right]
=\left[\begin{array}{cccc}\gamma_0&-\gamma_0&-\gamma_0&-q\gamma_0\\
-\gamma_1&\gamma_1&-\gamma_1&-q\gamma_1\\
-\gamma_2&-\gamma_2&\gamma_2&-q\gamma_2\\
q(2\gamma_0-\gamma)&q(2\gamma_1-\gamma)&
q(2\gamma_2-\gamma)&\gamma
\end{array}\right]\left[\begin{array}{c}\delta B_0/B_0\\
\delta B_1/B_1\\ \delta B_2/B_2\\ \delta\Phi\end{array}\right].
\end{equation}
The solution of such a linear differential equation is the exponential time
dependence $\exp(\lambda\tau)$ and the value of $\lambda$ is given by the 
characteristic equation
\begin{equation}
0=\lambda^4+a_1\lambda^3+a_2\lambda^2+a_3\lambda+a_4,
\end{equation}
where 
$$a_1=-2\gamma,\qquad a_2=\gamma^2(1+q^2)-4q^2(\gamma_0\gamma_1+
\gamma_0\gamma_2+\gamma_1\gamma_2),$$
$$a_3=4\gamma_0\gamma_1\gamma_2(1+3q^2),\qquad
a_4=-4(1+q^2)\gamma_0\gamma_1\gamma_2\gamma.$$
The equilibrium solution is stable if the real parts of all eigenvalues
$\lambda$ are negative. This is given by the Hurwitz criteria
$$W_1=a_1>0,\qquad W_2=a_1a_2-a_3>0,$$
$$W_3=a_3W_2-a_1^2a_4>0,\qquad W_4=a_4W_3>0.$$
It can be easily shown that only first and third criteria are independent in
our case. They are
\begin{equation}
\gamma=\gamma_0+\gamma_1+\gamma_2<0,
\end{equation}
\begin{equation}
4\gamma^3(1+q^2)-(1+3q^2)[4\gamma_0\gamma_1\gamma_2(1+3q^2)+2\gamma^3(1+q^2)-
8q^2\gamma(\gamma_0\gamma_1+\gamma_0\gamma_2+\gamma_1\gamma_2)]>0.
\end{equation}
The first criterion says that driving of the p-mode cannot be higher
than damping of the g-modes. The second criterion is a quadratic inequality
for $q^2$. It is fulfilled if $q^2$, and equivalently $|\Delta\sigma|$, 
is confined to a certain range or if it exceeds certain value.

If $\gamma_0\ll|\gamma_{1,2}|$, the equilibrium acoustic mode amplitude
(Eq.~12) is nearly equal the critical amplitude (Eq.~7). Moreover, according
to Wersinger \etal (1980), the terminal state is then a stable equilibrium
or a periodic limit cycle with average p-mode amplitude close to the
equilibrium value.
This means that the critical amplitude, given by Eq.~(7), may always be
treated as an estimate of the acoustic mode amplitude.

\subsubsection{Energy Equations}

In dynamical considerations it is often useful to use mode energies instead
of the amplitudes. The general formula for the energy of an oscillating mode
is $$E=\frac{1}{2}I\sigma^2|A|^2.$$
The energies in $1/2 I_0\sigma_0^2$ units in our case are
\begin{eqnarray}
E_0&=&|A_0|^2=B_0^2,\\
E_{1,2}&=&\frac{I_{1,2}\sigma_{1,2}^2}{I_0\sigma_0^2}|A_{1,2}|^2\approx
\frac{1}{2\epsilon}B_{1,2}^2,
\end{eqnarray}
where we used Eqs.~(4),(8) and we set
$$\sigma_{1,2}\approx\sigma_0/2,$$ because only the modes satisfying this
condition play a role.

Now we apply Eqs.~(9)--(11) to Eqs.~(19),(20) and we obtain
\begin{eqnarray}
\frac{{\rm d}E_0}{{\rm d}\tau}&=&2\gamma_0E_0+
2\epsilon C\sqrt{E_0E_1E_2}\sin\Phi,\\
\frac{{\rm d}E_{1,2}}{{\rm d}\tau}&=&2\gamma_{1,2}E_{1,2}-
C\sqrt{E_0E_1E_2}\sin\Phi,\\
\frac{{\rm d}\Phi}{{\rm d}\tau}&=&\Delta\sigma+C\left[\frac{\epsilon
\sqrt{E_1E_2}}{\sqrt{E_0}}-\frac{\sqrt{E_0}}{2}\left(\sqrt{\frac{E_1}{E_2}}+
\sqrt{\frac{E_2}{E_1}}\right)\right]\cos\Phi.
\end{eqnarray}
From these equations we see that the time derivative of the total energy
in oscillations is
\begin{equation}
\frac{{\rm d}E_{total}}{{\rm d}\tau}=\left\{
\begin{array}{lr}
2\gamma_0E_0+2\gamma_1E_1&\textrm{for\ the\ two-mode\ resonance}\\
2\gamma_0E_0+2\gamma_1E_1+2\gamma_2E_2&
\textrm{for\ the\ three-mode\ resonance}
\end{array}\right.
\end{equation}
The coupling terms are only responsible for the energy transfer
between the acoustic mode and the gravity modes. Since the growth rates
have different signs it is possible to have the solution with constant
total energy, \ie the equilibrium state, or with the total energy
varying around some average value, \ie the limit cycle.

\subsubsection{Equal Damping Rates}

We noted earlier that the modes in all important g-mode pairs have
approximately equal frequencies, $\sigma_1\approx\sigma_2$.
It turns out that their damping rates have the same property,
$\gamma_1\approx\gamma_2\equiv\gamma_g$.
If we assume exact equality we see from Eq.~(22) that
$$\frac{{\rm d}(E_1-E_2)}{{\rm d}\tau}=2\gamma_g(E_1-E_2).$$
Since $\gamma_g<0$, the difference between the two energies decays
exponentially. Thus, we further assume that the energies, as well as the real
amplitudes of the g-modes in each pair are equal. The energy of the
pair is $$E_g=E_1+E_2=2E_1$$ and then we simplify Eqs.~(21)--(23)
\begin{eqnarray}
\frac{{\rm d}E_0}{{\rm d}\tau}&=&2\gamma_0E_0+C\sqrt{E_0}E_g\sin\Phi,\\
\frac{{\rm d}E_g}{{\rm d}\tau}&=&2\gamma_gE_g-C\sqrt{E_0}E_g\sin\Phi,\\
\frac{{\rm d}\Phi}{{\rm d}\tau}&=&\Delta\sigma+C\left(\frac{E_g}
{2\sqrt{E_0}}-\sqrt{E_0}\right)\cos\Phi.
\end{eqnarray}
Identical equations are obtained in the two-mode resonance case, where
$\epsilon=1/2$ and $E_g=E_1$.

\Subsection{The Case of Many G-Mode Pairs}

The approach adopted by Dziembowski and Kr{\'o}likowska (1985) was to 
consider an ensemble of g-mode pairs that may be excited through the 
parametric resonance instability and to determine probability distribution
for the p-mode amplitude at onset of the instability. They assumed that 
the nonlinear interaction with first excited pair will halt the growth of the 
p-mode amplitude. This could be partially justified when the constant 
amplitude solution was stable. In the case of the ZAMS star model
they considered this was often the case. However, in her unpublished PhD 
thesis Kr{\'o}likowska showed later that in somewhat evolved objects the 
interaction with most likely excited g-modes cannot halt the amplitude growth.
Therefore, we have to consider the case of many g-mode pairs interacting with
the p-mode.

\subsubsection{Amplitude Equations}

The amplitude equations describing the case of many g-mode pairs
are obtained as a straightforward generalization of Eqs.~(1),(2)
\begin{eqnarray}
\frac{{\rm d} A_0}{{\rm d} \tau}&=&\gamma_0 A_0+
\sum_{j=1}^N{\rm i}\frac{\epsilon_j H_j}
{2\sigma_0 I_0}A_1^j A_2^j {\rm e}^{-{\rm i}\Delta\sigma_j\tau},\\
\frac{{\rm d} A_{1,2}^j}{{\rm d} \tau}&=&\gamma_{1,2}^j A_{1,2}^j+
{\rm i}\frac{H_j}{2\sigma_{1,2}^j I_{1,2}^j}
A_0 A_{2,1}^{j*} {\rm e}^{{\rm i}\Delta\sigma_j\tau},
\end{eqnarray}
where $N$ is the total number of pairs taken into account. Using the same
procedure as in the case of a single pair we obtain real amplitude equations
\begin{eqnarray}
\frac{{\rm d} B_0}{{\rm d} \tau}&=&\gamma_0 B_0+
\sum_{j=1}^N \frac{C_j}{2}B_1^j B_2^j \sin\Phi_j,\\
\frac{{\rm d} B_{1,2}^j}{{\rm d} \tau}&=&\gamma_{1,2}^j B_{1,2}^j-\frac{C_j}
{2}B_0 B_{2,1}^j \sin\Phi_j,\\
\frac{{\rm d} \Phi_j}{{\rm d} \tau}&=&\Delta\sigma_j+
\sum_{k=1}^N\frac{C_kB_1^k B_2^k}{2B_0}\cos\Phi_k-
\frac{C_j}{2}B_0\left(\frac{B_2^j}{B_1^j}+\frac{B_1^j}{B_2^j}\right)
\cos\Phi_j.
\end{eqnarray}

Now, let us try to find multimode stationary solution of the system. Putting
left-hand sides of Eqs.~(30)--(32) equal to zero we obtain
\begin{eqnarray}
\gamma_0B_0&=&-\sum_{j=1}^N \frac{C_j}{2}B_1^j B_2^j \sin\Phi_j,\\
\gamma_{1,2}^jB_{1,2}^j&=&\frac{C_j}{2}B_{2,1}^j B_0 \sin\Phi_j,\\
\frac{C_j}{2}B_0\left(\frac{B_2^j}{B_1^j}+\frac{B_1^j}{B_2^j}\right)\cos\Phi_j
&=&\Delta\sigma_j+\frac{1}{B_0}\sum_{k=1}^N \frac{C_k}{2}B_1^kB_2^k\cos\Phi_k.
\end{eqnarray}
From Eq.~(34) we get
\begin{equation}
\frac{C_j}{2}B_0\frac{B_{2,1}^j}{B_{1,2}^j}=\frac{\gamma_{1,2}^j}{\sin\Phi_j},
\end{equation}
and
\begin{equation}
B_1^jB_2^j=\frac{2\gamma_1^j(B_1^j)^2}{B_0C_j\sin\Phi_j}.
\end{equation}
From Eqs.~(35)--(37) we obtain
\begin{equation}
(\gamma_1^j+\gamma_2^j)\cot\Phi_j=\Delta\sigma_j+\frac{1}{B_0^2}
\sum_{k=1}^N \gamma_1^k(B_1^k)^2\cot\Phi_k.
\end{equation}
If we multiply Eq.~(34) for the mode $1$ by the same equation for the mode $2$
we get
\begin{equation}
\gamma_1^j\gamma_2^j=\frac{C_j^2}{4}B_0^2\frac{1}{1+\cot^2\Phi_j}.
\end{equation}
Eqs.~(38),(39) form a set of $2N$ equations for $N+2$ unknowns, namely $B_0^2$,
$$\sum_{k=1}^N\frac{\gamma_1^k(B_1^k)^2\cot\Phi_k}{C_k},$$ 
and $N$ values $\cot\Phi_j$.
This set has in general no solutions if the number of equations is larger than 
the number of unknowns. This happens for $N>2$. We see that
the stationary solution exists only if the number of excited pairs is 
not larger than $2$. In further sections we will study time evolution of 
the mode amplitudes $B$ and phases $\Phi$ by means of numerical integration of
Eqs.~(30)--(32).

\subsubsection{Statistical Equilibrium}

In the absence of stationary multimode solutions we may still expect that the
system will reach some kind of statistical equilibrium. In this section we
find a necessary condition for the existence of such a state. We restrict
our considerations to the case of equal damping rates in each pair,
$\gamma_1^j=\gamma_2^j\equiv\gamma_j$. As we mentioned earlier, this is
not far from the relevant realistic situation. Then, similarly as in the case
of one g-mode pair, we have $B_1^j=B_2^j\equiv B_j$ and
Eqs.~(30)--(32) become
\begin{eqnarray}
\frac{\rm{d} B_0}{\rm{d}\tau}&=&\gamma_0 B_0+\sum_{j=1}^N\frac{C_j}{2}
B_j^2\sin\Phi_j,\\
\frac{\rm{d} B_j}{\rm{d}\tau}&=&\gamma_j B_j-\frac{C_j}{2}B_0B_j\sin\Phi_j,\\
\frac{\rm{d} \Phi_j}{\rm{d}\tau}&=&\Delta\sigma_j+\sum_{k=1}^N
\frac{C_kB_k^2}{2B_0}\cos\Phi_k-C_j B_0 \cos\Phi_j.
\end{eqnarray}
Now, we introduce new variables
\begin{equation}
X=B_0^2,\quad Y_j=B_j\sin\Phi_j,\quad Z_j=B_j\cos\Phi_j.
\end{equation}
Then, from Eqs.~(40)--(42) we obtain
\begin{eqnarray}
\dot{X}&=&2\gamma_0 X+\sqrt{X}\sum_{k=1}^N C_k Y_k\sqrt{Y_k^2+Z_k^2},\\
\dot{Y}_j&=&\gamma_j Y_j-\frac{C_j\sqrt{X}}{\sqrt{Y_j^2+Z_j^2}}
\left(\frac{Y_j^2}{2}+Z_j^2\right)+\nonumber\\
& &+Z_j\left(\Delta\sigma_j+
\frac{1}{2\sqrt{X}}\sum_{k=1}^N C_k Z_k\sqrt{Y_k^2+Z_k^2}\right),\\
\dot{Z}_j&=&\gamma_j Z_j+\frac{C_j\sqrt{X}Y_j Z_j}{2\sqrt{Y_j^2+Z_j^2}}-
Y_j\left(\Delta\sigma_j+\frac{1}{2\sqrt{X}}
\sum_{k=1}^N C_k Z_k\sqrt{Y_k^2+Z_k^2}\right).
\end{eqnarray}
The divergence of the flow tells us whether the volume in the
phase space $(X,Y_j,Z_j)$ evolved by following the orbits obtained from 
Eqs.~(44)--(46) expands or shrinks exponentially. If it expands, the system
cannot reach the
statistical equilibrium. After some algebra we obtain the divergence of the
flow
\begin{equation}
\frac{\partial \dot{X}}{\partial X}+\sum_{j=1}^N
\left(\frac{\partial \dot{Y}_j}
{\partial Y_j}+\frac{\partial \dot{Z}_j}{\partial Z_j}\right)=
2(\gamma_0+\sum_{j=1}^N \gamma_j).
\end{equation}
From this we obtain the necessary condition for the stability of the
statistical equilibrium, which is
\begin{equation}
-\sum_{j=1}^N \gamma_j>\gamma_0,
\end{equation}
\ie the summed damping rate has to be larger than the acoustic mode
driving rate. This is the generalization of the criterion obtained by
Wersinger \etal (1980) in the case of a single g-mode pair.

In the same way as in the case of a single g-mode pair we derive
equations for the time evolution of energies of the interacting modes
\begin{eqnarray}
\frac{\rm{d} E_0}{\rm{d}\tau}&=&2\gamma_0 E_0+\sqrt{E_0}\sum_{j=1}^NC_j
E_j\sin\Phi_j,\\
\frac{\rm{d} E_j}{\rm{d}\tau}&=&2\gamma_j E_j-\sqrt{E_0}C_jE_j\sin\Phi_j,\\
\frac{\rm{d} \Phi_j}{\rm{d}\tau}&=&\Delta\sigma_j+\frac{1}{2\sqrt{E_0}}
\sum_{k=1}^NC_kE_k\cos\Phi_k-C_j\sqrt{E_0}\cos\Phi_j,
\end{eqnarray}
where $E_j$ is the energy of the $j$-th pair.
From Eqs.~(49),(50) we see that the time derivative of the total energy in
the system is given by
\begin{equation}
\frac{{\rm d}E_{\rm total}}{{\rm d}\tau}=\frac{{\rm d}E_0}{{\rm d}\tau}+
\sum_{j=1}^N \frac{{\rm d}E_j}{{\rm d}\tau}=
2\left(\gamma_0E_0+\sum_{j=1}^N\gamma_j E_j\right).
\end{equation}
If the damping rates are small compared to the
growth rate, the number $N$ of actively interacting g-modes necessary to halt
the acoustic mode growth is large, according to Eq.~(48). In such a case
the system reaches a state with time-dependent and limited amplitudes.
The average time derivative of the total energy (Eq.~52) is then equal to zero.

If we assume marginal stability, \ie the sum of damping rates
 of all active g-mode
pairs is just equal to the acoustic mode growth rate (see. Eq~48), then
the total energy derivative (Eq.~52) may be written as
\begin{equation}
\frac{{\rm d}E_{\rm total}}{{\rm d}\tau}=
2\left(\gamma_0 E_0+\bar{E}_j
\sum_{j=1}^N\gamma_j\right)=2\gamma_0\left( E_0-\bar{E}_j\right),
\end{equation}
where $\bar{E}_j$ is the mean energy of active g-mode pairs, weighted with
the damping rates. In the statistical equilibrium the time-averaged total
energy derivative is equal zero, which implies
\begin{equation}
\langle E_0\rangle=\langle\bar{E}_j\rangle.
\end{equation}
This equation looks like an energy equipartition. In fact,
the summed damping is higher than the driving (Eq.~48), so the mean
g-mode pair energy is lower than the acoustic mode energy.

One could think that Eq.~(54) could be used to estimate the acoustic
mode amplitude in the statistical equilibrium. It would be possible if we
assumed that active were only pairs which are unstable according to Eq.~(7)
with the acoustic mode amplitude averaged over the long time.
However, it has to be checked numerically whether such an assumption is
justified.

\Section{Time-Dependent Solutions}

In this section we analyze time evolution of simple systems. This will help us
to understand behavior of modes in realistic stellar model. We will see that
there is a qualitative similarity between simple cases and more realistic ones.

\Subsection{Single G-Mode Pair}

The case of a single pair of stable modes was studied in details by 
Wersinger \etal (1980). Here we present two examples of such system.

If the conditions for the stability of the equilibrium solution are satisfied
the system reaches this equilibrium. More interesting is the case of unstable
equilibrium and stable periodic limit cycle. An example is presented
in Fig.~1. 
\begin{figure}[!p]
\centerline{\includegraphics[width=1\textwidth]{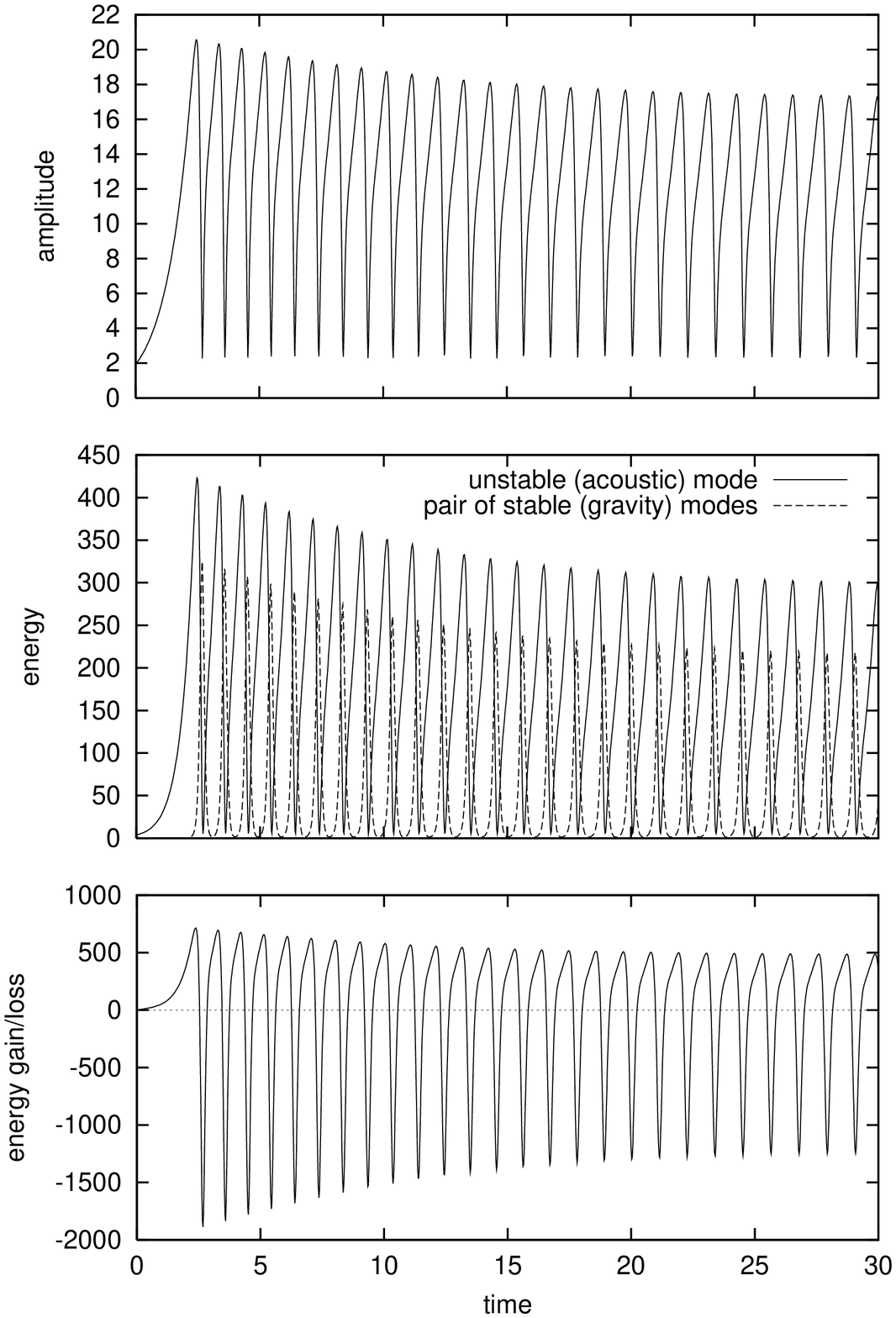}}
\FigCap{Time evolution of the system of an unstable (acoustic) mode and a pair
of stable (gravity)
modes. The parameters are $\gamma_0=1$, $\gamma_{1,2}=-3$, $\Delta\sigma=2$,
$C=1$. Upper panel shows the amplitude of the acoustic mode. Middle panel
shows energies of the acoustic mode (solid line) and the g-mode pair
(dashed line). Lower panel shows the time derivative of the total energy 
(Eq.~24). The value of $\gamma_0$ sets the unit of time and frequency.
The units of amplitudes and energies are set by the value of coupling
coefficient.}
\end{figure}

Initially, the amplitudes are low and the acoustic mode amplitude
grows according to the linear driving rate. When the amplitude becomes
sufficiently
high there is significant transfer of energy to the g-mode pair and
ultimately the acoustic mode amplitude growth is halted. The evolution
proceeds rapidly on the coupling timescale. First, nearly all energy of the
p-mode is transfered to the g-mode pair. Then the energy is returned back to
the p-mode, but not completely. Part of the energy is dissipated.
The state of constant amplitude modulation
is reached only after the excess of energy gained in the initial phase is
dissipated, \ie after the time about $20/\gamma_0$. This excess is an
artifact of the initial conditions. The mean energy balance (Eq.~52) implies
\begin{equation}
\frac{\langle\bar{E_g}\rangle}{\langle E_0\rangle}=
\frac{\gamma_0}{|\gamma_g|}.
\end{equation}

The period of the limit cycle is about $1/\gamma_0$. It is so, because
the system spends most of the time in the phase of exponential growth
of the acoustic mode amplitude. Finally, from the bottom panel we see that
the averaged time derivative of the total energy is indeed equal zero.

Fig.~2 presents the case
when neither a stable equilibrium nor a stable limit cycle exist. Such a
situation takes place,
in particular, when condition given by Eq.~(48) with $N=1$ is not satisfied.
The amplitude growth is unlimited. This growth is modulated in a 
similar way as in the periodic limit cycle solution, but amplitudes
in consecutive cycles are higher and higher. The reason why the
situation is unstable is that the g-mode damping is low and the 
pair does not manage to loose enough energy to balance the unstable mode
driving. In lower panel of Fig.~2 it is clearly seen that the average
time derivative of the total energy is positive.
\begin{figure}[!p]
\centerline{\includegraphics[width=1\textwidth]{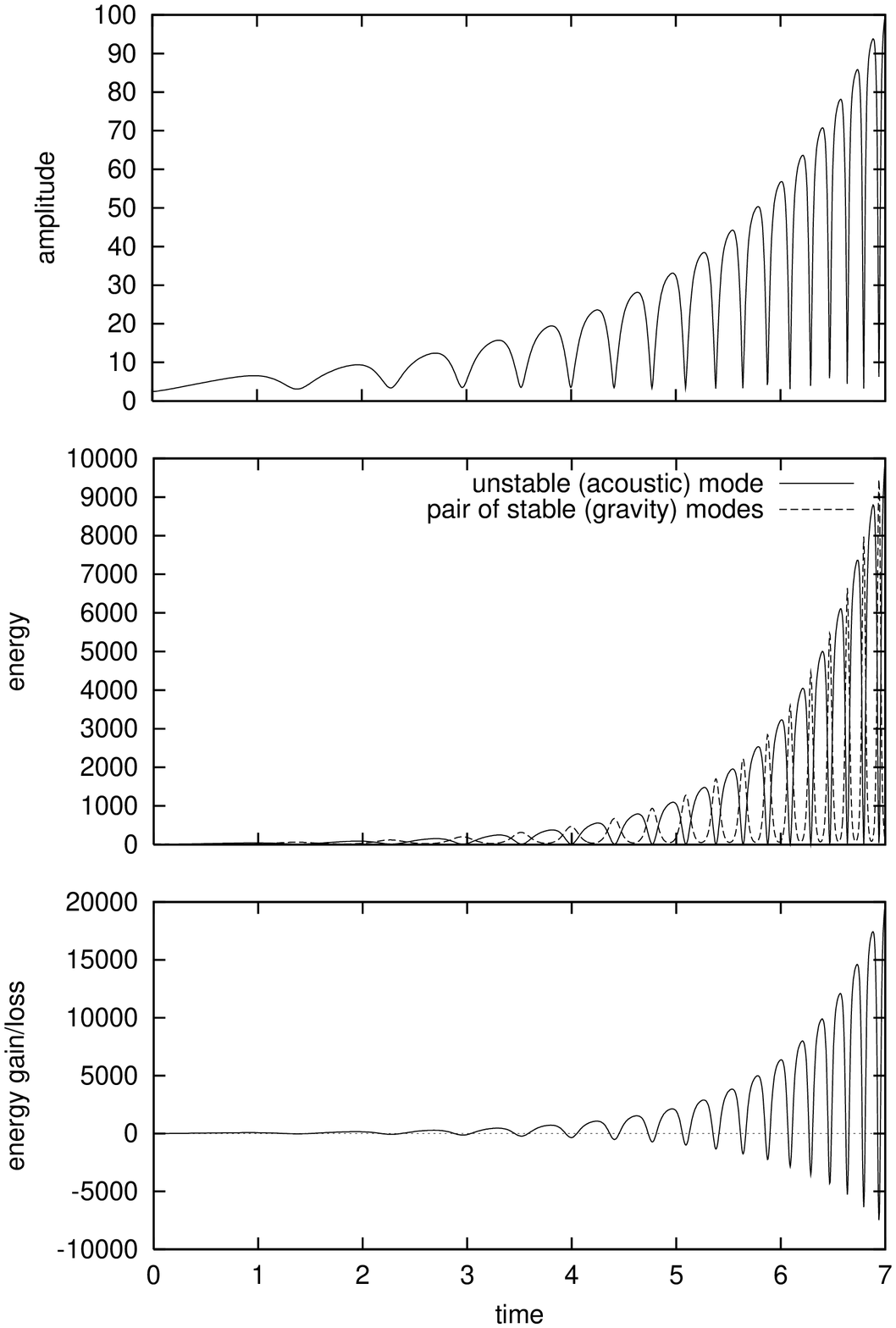}}
\FigCap{The same as in Fig.~1 but with parameters that exclude stable
solutions, either equilibrium, or limit cycle.
The parameters are $\gamma_0=1$, $\gamma_{1,2}=-0.4$, $\Delta\sigma=2$,
$C=1$.} 
\end{figure}

We have to add here, that Eq.~(48) is not a sufficient condition for
the stability. Wersinger \etal (1980) showed examples of solutions with
unlimited growth despite satisfying this condition for $N=1$ (a single
g-mode pair). We found similar cases for $N>1$. What is needed to stop
the amplitude growth is some excess of the damping.

\Subsection{Many G-Mode Pairs}

In this section we study the time evolution of a simplified system of
interacting modes in the case when several pairs are necessary to halt the 
acoustic mode growth. We set, as previously, $\gamma_0=1$ and $C_j=1$ for
all $N=19$ pairs.
The damping rates are a few times smaller than the growth rate and for all
g-modes we set $\gamma_j=-0.3$. The detuning parameters, $\Delta\sigma_j$,
are equally spaced between $-13.824$ and $13.824$ with the separation $1.536$.
Energies of all interacting modes are shown in Fig.~3 and
the time derivative of the total energy is shown in Fig.~4.
\begin{figure}[!ht]
\centerline{\includegraphics[angle=270,width=1\textwidth]{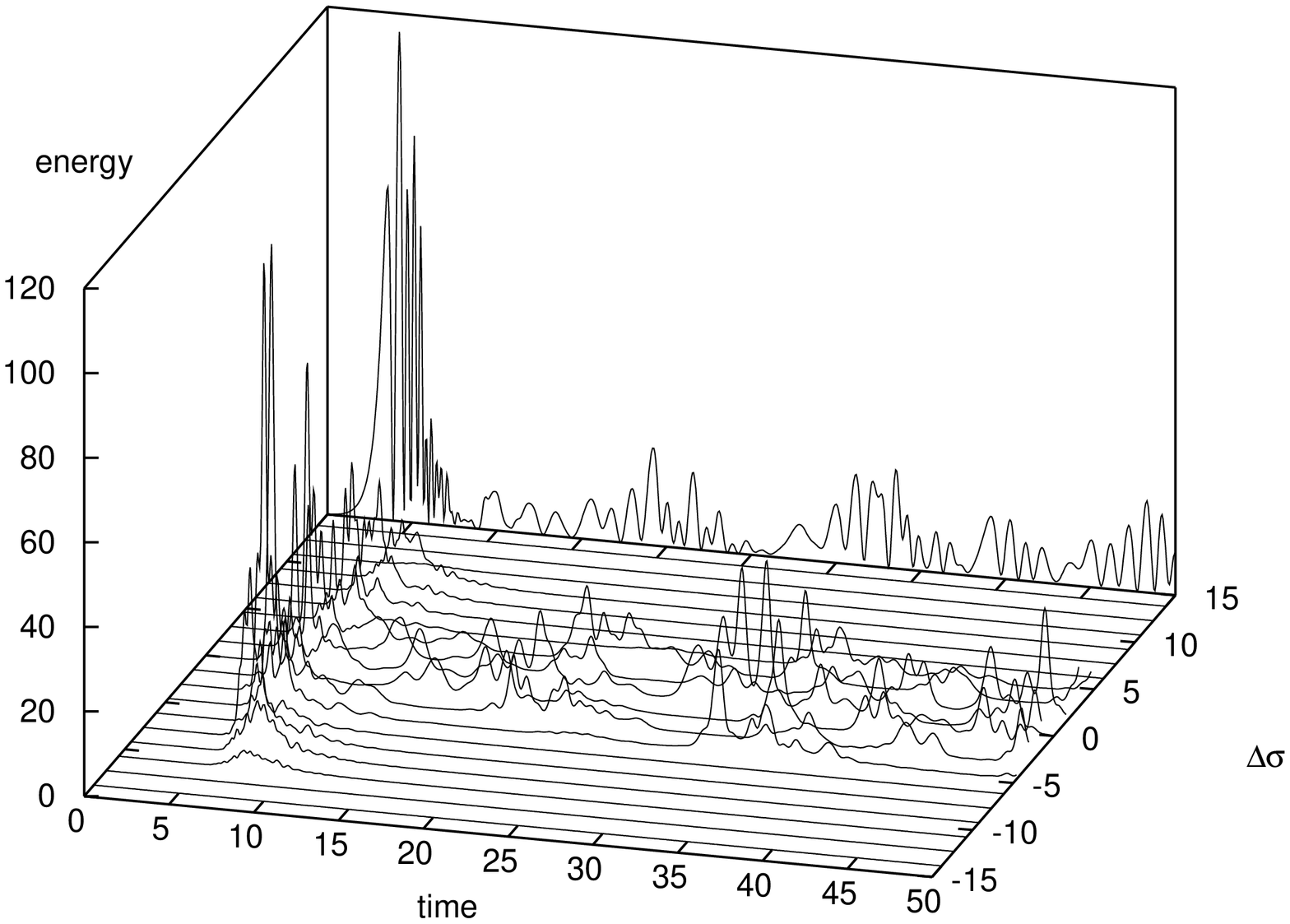}}
\FigCap{The time evolution of the multimode system. The energy of the
acoustic mode is plotted in the back wall of the 3-D plot. The remaining
lines denote energies of g-mode pairs. The detuning parameters of the pairs
are equally spaced and the difference between consecutive $\Delta\sigma$'s
is $1.536$. Other parameters are the same for all pairs, \ie
$\gamma_j=-0.3$, $C_j=1$. The growth rate of the acoustic mode is
$\gamma_0=1$. This sets the units of time and frequency. The unit of energy
is determined by the value of coupling coefficients. Initial energies of the
g-mode pairs are set randomly around the value of $0.01$. Initial acoustic
mode energy is $0.3$.}
\end{figure}
\begin{figure}[!ht]
\centerline{\includegraphics[angle=270,width=1\textwidth]{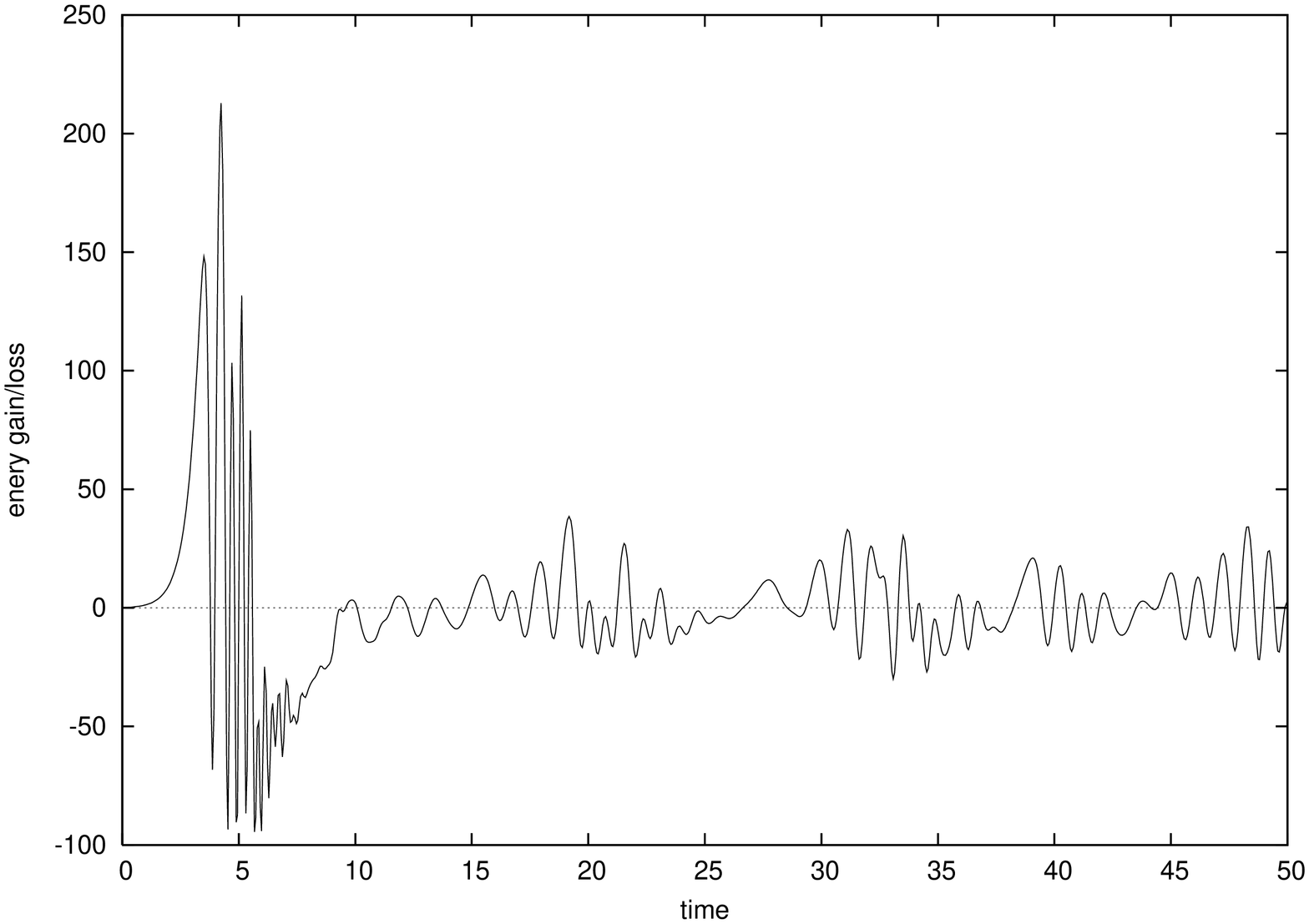}}
\FigCap{The time derivative of the total energy in the system shown in
Fig.~3.}
\end{figure}

At the beginning the acoustic mode amplitude grows exponentially.
When the amplitude reaches approximately the critical value given in Eq.~(7)
the energy transfer begins.
The lowest critical amplitude among the pairs in our system is for the one
with $\Delta\sigma=0$ and this pair is excited first. However,
this pair alone cannot halt the p-mode growth since the damping rate is too
low. Gradually, other pairs are excited, but it takes some time before their
amplitudes are large enough to affect the p-mode growth. The initial
excess energy is dissipated by modes in a rather wide range of detuning
parameters.

Ultimately, after the time of about $15/\gamma_0$ the system achieves
the statistical equilibrium in which only $7$ closest-to-resonance
pairs survive. Let us note that the number implied by the necessary stability
condition (Eq.~48) is 4.
The rapid oscillations seen in this phase occur on the timescale determined
by the largest detuning parameter of the active pairs and this timescale
turns out to be crudely equal to the nonadiabatic timescale $\gamma_0^{-1}$.
Surprisingly, the result is similar as in the case of a single g-mode pair.
This result, which is essential for understanding of amplitude limitation
mechanism will be explained in Section~4.

The overall
behavior of the acoustic mode is characterized by an irregular variability
with a large amplitude modulations. In Fig.~4 we see that the total energy of
oscillations is conserved only on average, which means that there is energy
exchange with the background.

Fig.~5 presents the
energies of active pairs (upper panel) and the energy budget for the
acoustic mode (lower panel) in a selected time-interval in the statistical
equilibrium phase. What is important, the local energy maxima of various
g-mode pairs occur at various times. This lack of synchronization is essential.
Synchronized pairs act just as a single pair, as we will show in the next
subsection, and they are not efficient energy receivers.

In the lower panel we see that the acoustic mode energy
balance is zero on average. The coupling causes an appreciable energy
exchange between the acoustic and the gravity modes. The net energy transfer
necessary to keep the g-modes alive is relatively small.
\begin{figure}[!ht]
\centerline{\includegraphics[angle=270,width=1\textwidth]{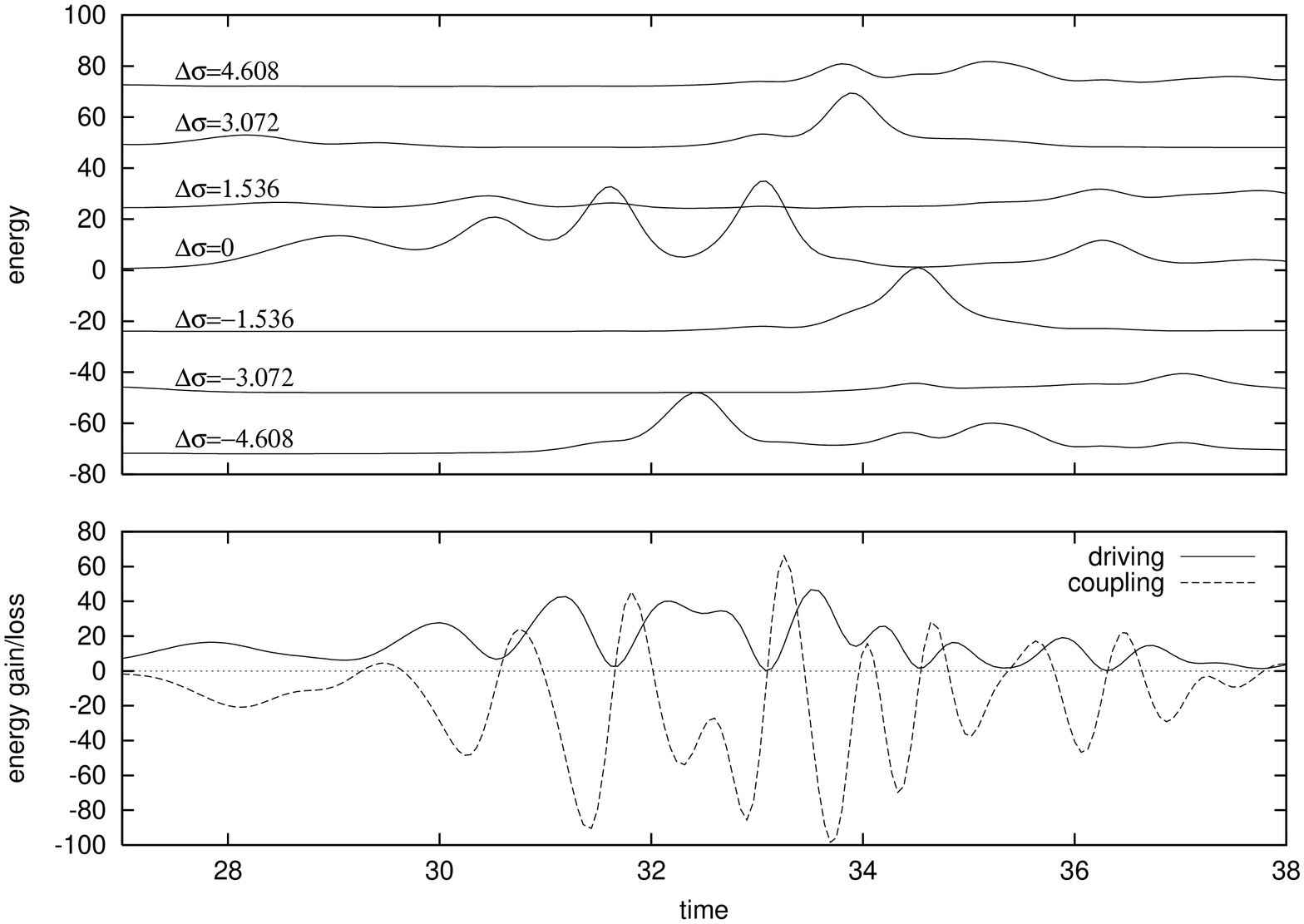}}
\FigCap{Upper panel: energies of active pairs in the statistical equilibrium;
the zero-levels of the energies are shifted proportionally to $\Delta\sigma$.
Lower panel: driving of the acoustic mode (solid line) given by the first term
in Eq.~(49) and energy transfer to the g-mode pairs due to coupling
(dashed line), given by the sum in Eq.~(49).}
\end{figure}

The rapid and nearly complete energy exchange between the acoustic and the
gravity modes,
manifested by very large range of amplitude variations, is similar to that in
the case of a single g-mode pair. This is so because each individual energy
exchange act is typically controlled by a single pair. The irregularity
in the present case is caused by involvement of other pairs. In Fig.~5 we see
examples when a new pair takes over the control. This happens when the
previous pair have low amplitude. Which of the remaining pairs takes the
control depends on the relative phase. It must be favorable, \ie
$|A_0|\sin\Phi_j$ must be large negative.

From energy balance (Eq.~52) we obtain
\begin{equation}
\frac{\langle\bar{E_j}\rangle}{\langle E_0\rangle}=
\frac{\gamma_0}{\sum_{j=1}^N|\gamma_j|},
\end{equation}
where the summation is done over the active pairs.
This is essentially the same result as in the case of a single g-mode pair
(see Eq.~55). In our example, the sum of the damping rates of the active
pairs is about $2$ times higher than the growth rate, $\gamma_0$. 
Thus, the average g-mode pair energy is
about $2$ times lower than the average energy of the p-mode.

It is of interest that most of the active g-mode pairs should not be excited
according to the parametric instability criterion (Eq.~7) applied to the
time-averaged p-mode amplitude.
However, the formula for the critical amplitude is valid only
when the p-mode complex amplitude, $A_0$, is assumed constant or
to vary sufficiently slowly. Here this assumption is not satisfied. One
can easily see how the varying $A_0$ modifies the parametric excitation
criterion.

Let us assume $A_0=A_0'{\rm e}^{{\rm i}\eta\tau}$, and
$A_0'$ is constant. Then, Eq.~(2) becomes
\begin{equation}
\frac{{\rm d}A_{1,2}}{{\rm d}\tau}=\gamma_{1,2}A_{1,2}+{\rm i}
\frac{H}{2\sigma_{1,2}I_{1,2}}A_0'A_{2,1}^*
{\rm e}^{{\rm i}(\Delta\sigma+\eta)\tau},
\end{equation}
\ie only detuning parameter is modified. This means the criterion in Eq.~(7)
becomes
\begin{equation}
|A_0'|=|A_0|>B_{cr}'\equiv\sqrt{\frac{4\gamma_1\gamma_2}{C^2}\left[1+\left(
\frac{\Delta\sigma+\eta}{\gamma_1+\gamma_2}\right)^2\right]}.
\end{equation}
The effective detuning parameter is now $\Delta\sigma+\eta$.
In particular, if $\eta=-\Delta\sigma$ the criterion becomes much less
restrictive for the pairs with detuning parameter $\Delta\sigma$ than 
in the case of $A_0=const$.

In the case presented in Figs.~(3)--(5) the amplitude of the acoustic
mode varies in a very complicated way. We may treat these variations
as consisting of many components oscillating with various frequencies $\eta$.
A g-mode pair with the detuning parameter $\Delta\sigma_j$ is affected mainly
by components with $\eta\approx-\Delta\sigma_j$. This implies that pairs with
relatively
large detuning parameters may be excited even when the average acoustic mode
amplitude is below the critical value given by Eq.~(7). Therefore, we
cannot estimate average acoustic mode amplitude in a simple way described
at the end of Section~2.

\Subsection{Pairs with Close Detuning Parameters}

When we conducted similar numerical experiments with two or more g-mode pairs
having very close detuning parameters, $\Delta\sigma_j$, we observed a
synchronization of phases of these pairs.
This effect may be demonstrated analytically with the use of Eqs.~(40)--(42).

Let us consider two pairs, $j_1$, and $j_2$ with
$\Delta\sigma_{j_1}=\Delta\sigma_{j_2}$, as well as
$\gamma_{j_1}=\gamma_{j_2}\equiv\gamma$
and $C_{j_1}=C_{j_2}\equiv C$. Then, from Eq.~(42), we have
\begin{equation}
\frac{\rm d}{{\rm d}\tau}\frac{\Phi_{j_1}-\Phi_{j_2}}{2}=C B_0 
\frac{\cos\Phi_{j_2}-\cos\Phi_{j_1}}{2}=
S\sin\frac{\Phi_{j_1}-\Phi_{j_2}}{2},
\end{equation}
where
\begin{equation}
S\equiv CB_0\sin\frac{\Phi_{j_1}+\Phi_{j_2}}{2}.
\end{equation}
Upon integration we get
\begin{equation}
\tan\frac{\Phi_{j_1}-\Phi_{j_2}}{4}=
\tan\frac{\left(\Phi_{j_1}-\Phi_{j_2}\right)_0}{4}
\exp\left(\int S{\rm d}\tau'\right).
\end{equation}
The integrand is a rapidly varying function, but it has a nonzero mean value.
This value must be less than zero. To see this let us write Eq.~(41) for
the two pairs in the form
\begin{equation}
\frac{{\rm d}\ln B_{j_{1,2}}}{{\rm d}\tau}=\gamma-
\frac{C}{2}B_0\sin\Phi_{j_{1,2}}.
\end{equation}
In the statistical equilibrium we have
\begin{equation}
C\langle B_0\sin\Phi_{j_{1,2}}\rangle=2\gamma<0.
\end{equation}
This holds if both phases $\Phi_{j_{1,2}}$ are in the $(-\pi,0)$ range when
$B_0$ is large.
Then, the average of the two phses, $(\Phi_{j_1}+\Phi_{j_2})/2$, is also in
this range during large $B_0$ phase, implying
\begin{equation}
\langle S\rangle<0.
\end{equation}
In fact, the values of the sine functions of both phases as well as of their
average, are of the same order, except
short periods of time when the phases are close to $0,\pm\pi$. This implies
the left-hand side of Eq.~(64) to be of the order of $2\gamma$,
like in Eq.~(63), and the synchronization occurs on the timescale
$|\gamma|^{-1}$.

The phase synchronization has an important
consequence for the multimode coupling. We may replace in Eqs.~(49)--(51)
the energies of the pairs with the same $\Phi$, $\gamma$, $C$, and
$\Delta\sigma$ by the sum of their energies. Then, the equations for the other
pairs remain unchanged. This means that the synchronized pairs act
effectively as a single pair and in this way we may
reduce the number of pairs and the number of equations to solve.

In a more realistic situation, the two pairs do not have exactly equal
detuning parameters. Then, instead of Eq.~(59) we have
\begin{equation}
\frac{\rm d}{{\rm d}\tau}\frac{\Phi_{j_1}-\Phi_{j_2}}{2}=
\frac{\Delta\sigma_{j_1}-\Delta\sigma_{j_2}}{2}+
S\sin\frac{\Phi_{j_1}-\Phi_{j_2}}{2}.
\end{equation}
When we use Eq.~(64), then Eq.~(65) becomes
\begin{equation}
\frac{\rm d}{{\rm d}\tau}\frac{\Phi_{j_1}-\Phi_{j_2}}{2}\approx
\frac{\Delta\sigma_{j_1}-\Delta\sigma_{j_2}}{2}+
2\gamma\sin\frac{\Phi_{j_1}-\Phi_{j_2}}{2}.
\end{equation}
It's solution is
\begin{equation}
\tan\frac{\Phi_{j_1}-\Phi_{j_2}}{4}=
\frac{1}{z}-
\frac{1}{z}\sqrt{1-z^2}\tanh\left(\sqrt{1-z^2}|\gamma|\tau+T\right),
\end{equation}
where $z=(\Delta\sigma_{j_1}-\Delta\sigma_{j_2})/(4|\gamma|)$, and $T$ is
the integration constant whose value is determined by the initial conditions.
If $|z|\ll1$ then for $\tau\rightarrow\infty$ we get
\begin{equation}
\Phi_{j_1}-\Phi_{j_2}\,\approx\,2z\,=\,
\frac{\Delta\sigma_{j_1}-\Delta\sigma_{j_2}}{2|\gamma|}.
\end{equation}
We see now, that the phases are synchronized if the
difference of detuning parameters is much less than $|\gamma|$.
In fact, we found in numerical experiments that the pairs
may be treated as synchronized if the $\Delta\sigma$ difference is just
smaller than $\gamma$. Only for larger difference the pairs are really
independent.

The synchronization has important consequence for the stability condition
(Eq.~48) in which we should count only independent pairs.

\Section{General Properties of Multimode Solutions}

In the previous section we studied systems where all pairs had the same
damping rates and coupling coefficients. It turns out, however, that
the main qualitative properties of the multimode solutions are valid also
in more realistic systems, where the pairs have various damping rates
and coupling coefficients.

Firstly, the pairs are excited in significantly wider range of $\Delta\sigma$
than implied by Eq.~(7) with the acoustic mode amplitude replaced by it's
average value. Secondly, the pairs with very close detuning parameters
are synchronized. Thirdly, the p-mode amplitude is very strongly modulated
due to nearly complete energy transfer between the interacting modes.
Finally, the modulation timescale is given
by the inverse of the growth rate $\gamma_0$.

The wide range of detuning parameters of active pairs is the result of
rapidly varying amplitude of the acoustic mode (see Subsection~3.2.2).

The pairs with close detuning parameters and equal damping rates and coupling
coefficients are synchronized, as was shown in Subsection~3.3. It turns out
that the pairs with different damping
rates and/or coupling coefficients also tend to synchronization if their
detuning parameters differ less than the smaller of their damping rates.
In such a case, in a statistical equilibrium, we have
\begin{eqnarray}
\langle B_0\sin\Phi_{j_1}\rangle&=&\frac{2\gamma_{j_1}}{C_{j_1}},\\
\langle B_0\sin\Phi_{j_2}\rangle&=&\frac{2\gamma_{j_2}}{C_{j_2}},
\end{eqnarray}
(see Eq.~63). The synchronization causes both
left-hand sides of above equations to be roughly equal. 
If the right-hand
sides differ significantly, it is obvious that these equations cannot be
satisfied simultaneously. Only the one with lower value of $|\gamma/C|$
can be satisfied, because the sine functions have limited values.
The energy transfer to the g-mode pair with higher $|\gamma/C|$ is
inefficient and this pair gets damped. This means that we may treat such
pairs as one effective pair, like in the case of equal damping rates
and coupling coefficients. Here, this effective pair is just the one
with the smallest $|\gamma/C|$ ratio.

The large amplitude modulation is the result of nearly complete energy
exchange between the p-mode and the g-mode pair that currently controls the
interaction. The mode energies averaged over all active pairs and over 
a long time satisfy Eq.~(56). However, if we have groups of synchronized
pairs, each group has to be counted as a single pair.
We expect, just as in the simplified case considered in the previous section,
that the average energy of the g-modes is a few times lower than
the average p-mode energy.

The fact that the amplitude modulation timescale is of the order of
$\gamma_0^{-1}$
is caused by the synchronization of the g-mode pairs with close detuning
parameters and the stability criterion given by Eq.~(48).
In a realistic situation the damping rates
of the active g-mode pairs are much smaller than the driving rate of the
acoustic mode. This means that the number of active pairs is high, namely
it has to be higher than $N_{min}=\gamma_0/|\bar{\gamma_j}|$ (see Eq.~48),
where $\bar{\gamma_j}$ is the mean value of the damping rate of the
active pairs. Moreover, the independent active pairs have to be separated
in the $\Delta\sigma$ space by more than $|\bar{\gamma_j}|$. This implies
the range of detuning parameters occupied by active pairs have to be
larger than $N_{min}|\bar{\gamma_j}|=\gamma_0$. The maximum $\Delta\sigma$
of the active pairs, which gives the shortest modulation timescale, is thus
higher than $\gamma_0/2$. In fact, it should be typically equal to a few times
$\gamma_0$.

\Section{Application to a Realistic Stellar Model}

\Subsection{The Model}

Numerical study of oscillation in a realistic situation must begin with model
construction. In our work 
we adopted one of possible models of the $\delta$~Scuti star XX~Pyxidis 
calculated by Pamyatnykh \etal (1998). Our choice is  motivated, in part, by 
the interesting data that we have for this object and, in part, by 
the fact that the star is a moderately evolved object, just like most of the 
well studied $\delta$ Scuti stars.

The model mass is $1.9 M_\odot$, radius $2.06 R_\odot$, effective temperature
$8045 K$, and luminosity $15.8 L_\odot$. The model is slightly evolved,
the central hydrogen abundance is $0.39$. No rotation is assumed.

We computed nonadiabatic acoustic modes in our model. The unstable modes of
$l\le 2$ are found in the frequency range $3.7\lesssim\sigma\lesssim6.2$.
The most important parameters of the modes are listed in Table~1.
\MakeTable{|c|c|c|c|}{\textwidth}{Parameters of unstable acoustic modes}
{\hline
$l$ & $\sigma$ & $P$[min] & $\gamma$\\
\hline
0 & 3.909 & 52.91 & 1.455e-05\\
0 & 4.544 & 45.52 & 1.096e-04\\
0 & 5.201 & 39.77 & 3.527e-04\\
0 & 5.859 & 35.30 & 4.869e-04\\
\hline
1 & 4.125 & 50.15 & 3.758e-05\\
1 & 4.815 & 42.96 & 2.005e-04\\
1 & 5.495 & 37.64 & 4.812e-04\\
1 & 6.162 & 33.57 & 2.194e-04\\
\hline
2 & 3.725 & 55.53 & 3.041e-06\\
2 & 3.883 & 53.27 & 4.134e-06\\
2 & 4.440 & 46.59 & 8.830e-05\\
2 & 5.120 & 40.40 & 3.243e-04\\
2 & 5.793 & 35.71 & 5.106e-04\\
\hline}
Additionally, the growth rates of the acoustic modes versus their
frequencies are shown in Fig.~6.
\begin{figure}[!t]
\centerline{\includegraphics[angle=270,width=1\textwidth]{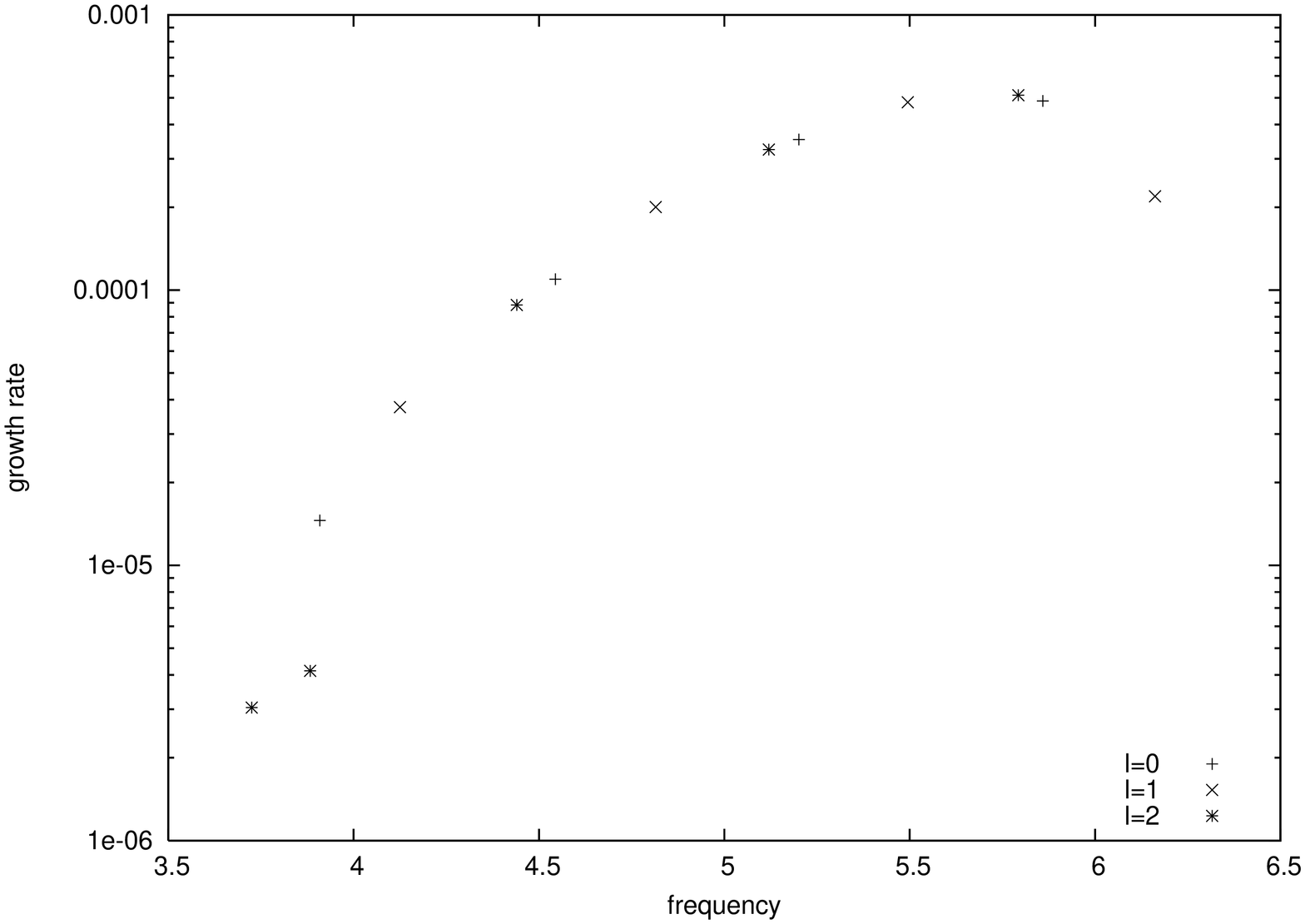}}
\FigCap{Dimensionless growth rates of the unstable acoustic modes of low-$l$
in our model versus their dimensionless frequencies.}
\end{figure}
We will also use
another linear characteristics of the modes, the complex factor $f$,
which is the ratio of relative flux to relative radius variations.
The absolute values of this factor are close to $30$ for all the modes listed
in Table~1.

\Subsection{Gravity Modes}

The properties of g-modes are determined by the 
behavior of the Brunt-V\"ais\"al\"a frequency. It is presented in Fig.~7.
Here we summarize the more general discussion presented by Dziembowski (1982)
and Dziembowski and Kr{\'o}likowska (1985).
\begin{figure}[htb]
\centerline{\includegraphics[width=0.9\textwidth]{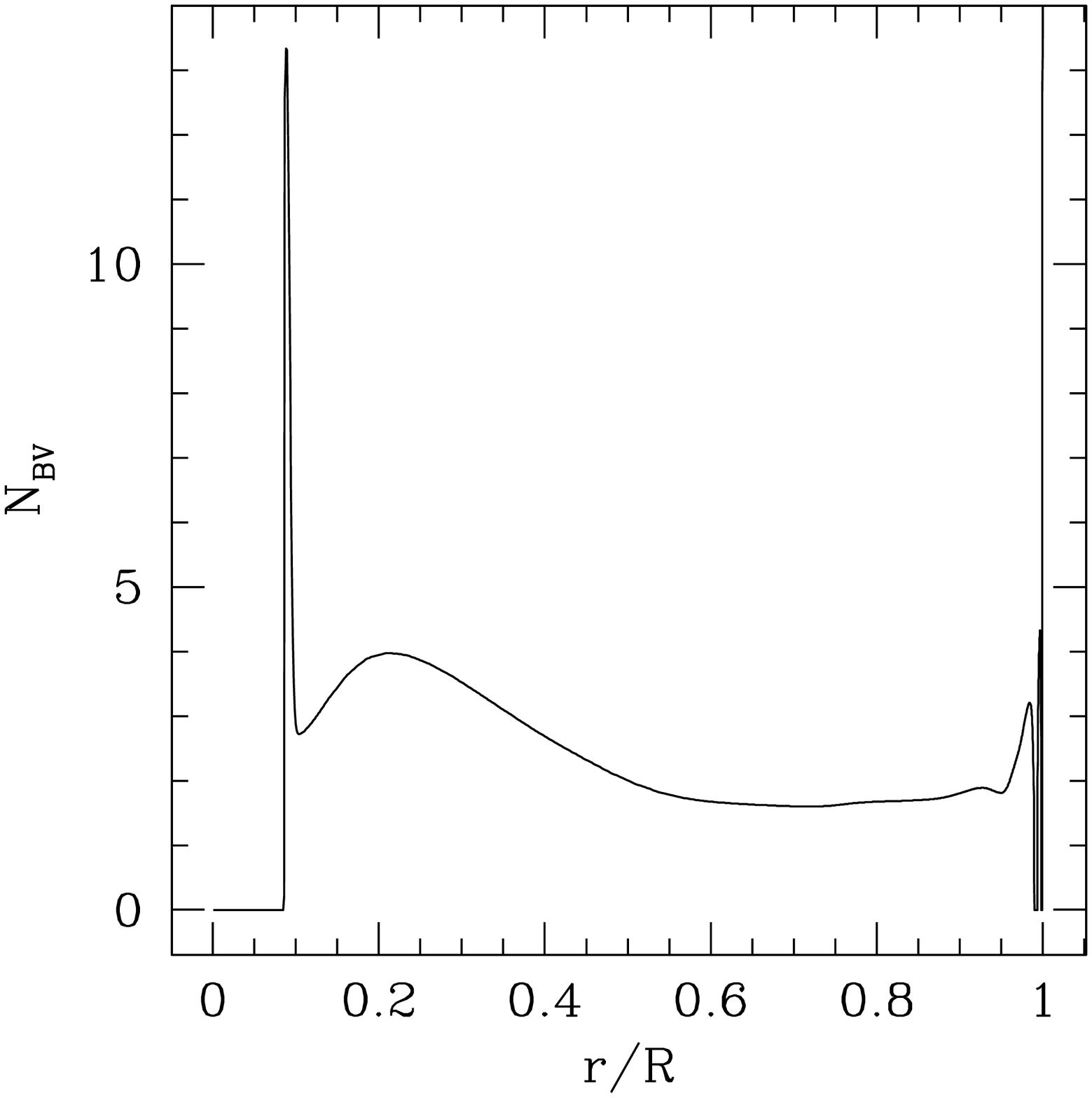}}
\FigCap{The dimensionless Brunt-V\"ais\"al\"a frequency versus radius in the
adopted model. The atmospheric values are about 30.} 
\end{figure}

Gravity modes can propagate in a region where their frequencies are lower than
the Brunt-V\"ais\"al\"a frequency (g-mode propagation zone) and this region 
have to be surrounded by evanescent zones where the Brunt-V\"ais\"al\"a
frequency is lower. The frequencies of the g-modes that are efficiently 
coupled to the acoustic mode are close to one half of its frequency. Thus,
as we may see in Table~1, the interesting
gravity modes have frequencies between about $1.8$ and $3.1$.
We see in Fig.~7 that there are two separate propagation zones for such
frequencies, one extending from $r/R\approx0.1$ to $r/R\approx0.4\div0.5$, and
another in the range $0.95\lesssim r/R\lesssim0.98$\footnote{In fact there is
another, very narrow, propagation zone at $r/R<0.1$
for modes with $\sigma\gtrsim 3$, but this has no practical
meaning.}. Therefore, we have to
consider two sets of g-modes of vastly different properties.

The inner g-modes can be studied in the asymptotic approximation (see, \eg
Dziembowski 1982, Van~Hoolst \etal 1998). It is valid when the wavelength of
the mode is much shorter than the local hightscale. The radial wavenumber of
a mode $k$ is given by
\begin{equation}
k_k=\frac{\kappa_k}{r}=\frac{1}{r}
\sqrt{\Lambda_k\left(\frac{{\mathcal N}_{BV}^2}{\sigma_k^2}-1\right)},
\end{equation}
where ${\mathcal N}_{BV}$ is the dimensionless Brunt-V\"ais\"al\"a frequency,
and $\Lambda_k=l_k(l_k+1)$.
The displacement, ${\boldsymbol \eta}_k$, is expressed in the form
\begin{equation}
{\boldsymbol\eta}_k({\boldsymbol r}, \tau)={\boldsymbol\xi}_k({\boldsymbol r})
A_k {\rm e}^{{\rm i}\sigma_k\tau},
\end{equation}
where the eigenvector
\begin{equation}
{\boldsymbol\xi}_k({\boldsymbol r})=
\left(y_{1,k}(r){\boldsymbol e}_r+z_k(r){\boldsymbol\nabla}_H\right)
Y_{l_k}^{m_k}(\theta,\phi).
\end{equation}
The radial eigenfunctions in the asymptotic approximation are given by
\begin{eqnarray}
y_{1,k}(r)&=&\frac{f_k(r)}{r^3\sqrt{\rho}},\\
z_k(r)&=&\frac{1}{\Lambda_kr^3\sqrt{\rho}}\left(\frac{{\rm d}f_k(r)}
{{\rm d}\ln r}+f_k(r)\frac{\mathcal{A}-V_g}{2}\right),
\end{eqnarray}
where
\begin{equation}
f_k(r)=\mathcal{D}_k\sqrt{r/\kappa_k}\sin\psi_k,\qquad\psi_k=
\int\kappa_k{\rm d}\ln r,
\end{equation}
the structure functions $\mathcal{A}$ and $V_g$ are given by
\begin{equation}
V_g=\frac{V}{\Gamma}=-\frac{1}{\Gamma}\frac{{\rm d}\ln p}{{\rm d}\ln r},\qquad
\mathcal{A}=-\frac{{\rm d}\ln\rho}{{\rm d}\ln r}-V_g,
\end{equation}
$\Gamma$ is the adiabatic exponent and remaining symbols have traditional
meaning. When the
mode degree $l_k$ is high so is the radial wavenumber,
implying high radial order, $n_k$. 
In such a case a small change of the frequency is sufficient to
change the order $n$ by one, which means the spectral density of g-modes is 
very high. Thus, for each $l$ it is quite easy to find a pair satisfying the
resonant condition. We find in our model that the frequency distance
between inner g-modes of a given $l$ and consecutive $n$'s, for the modes
that are interesting for us, can be approximated by
\begin{equation}
\delta\sigma_{l,n}\equiv\sigma_{l,n}-\sigma_{l,n+1}\approx
\frac{0.36\,\sigma_{l,n}}{l}.
\end{equation}

The inner g-modes are quasiadiabatic which means
the only important nonadiabatic quantity is the
damping rate. In this approach it may be approximated by
\begin{equation}
-\gamma_k=\frac{\Lambda_k}{\sigma_k^2\tau_{th}},
\end{equation}
where $\tau_{th}$ is the thermal timescale of the inner g-mode cavity. In our
case it is roughly equal to $10^{10}$, which implies the damping rates of the
relevant modes range from $3\times10^{-8}$ at $l\sim40$ to
$2\times10^{-5}$ at $l\sim1000$. Thus, the stability
criterion given by Eq.~(17) is not satisfied, because the driving rates
of the acoustic modes, given in Table~1, are typically much higher. It is
necessary to take into account at least several g-mode pairs to satisfy
Eq.~(48). As was shown in Subsection~2.2.1, this implies that the coupling
to the inner g-modes cannot result in constant amplitude p-mode pulsation.

Neither asymptotic nor quasiadiabatic approximation can be applied to modes
trapped in the outer g-mode cavity. These outer g-modes still have relatively
high degrees $l$ but,
because of the narrowness of the cavity, the radial orders $n$ are low,
typically $1$ or $2$. The nonadiabatic damping rates of these modes
are typically between about $0.02$ at $\sigma_{1,2}\lesssim2.1$ and 
about $0.5$ at $\sigma_{1,2}\gtrsim2.8$. The strong damping of these modes
might suggest they are not likely to be excited.
However, the coupling coefficients are also
orders of magnitude larger than in the case of the inner g-modes,
and the critical amplitudes given by Eq.~(7) can
be comparable to those for the latter. Neither in this case we expect
a stationary pulsation because the second of the stability criteria (Eq.~18)
is not fulfilled because he have $q^2\ll1$ and $\gamma_0\ll|\gamma_{1,2}|$.
Since the criteria given by Eq.~(17) as well as Eq.~(48) with $N=1$ are
satisfied and the damping rates are much higher than the growth rates,
a single outer g-mode pair is sufficient to halt the acoustic
mode growth. The instability of the stationary solution in this case implies
that the system reaches a limit cycle solution.

In cooler $\delta$~Scuti stars p-mode driving extends to lower frequencies,
such that the resonant g-modes have their frequencies below the 
$\mathcal N_{BV}$ minimum. In this case the g-modes propagate through the
whole radiative interior. Clearly the spectrum of such modes is very dense
and damping is intermediate between that of the inner and the outer g-modes.

\Subsection{The Coupling Coefficient}

Various authors give various expressions for
the coupling coefficient. While the form of equations is basically always
the same, the coupling coefficient may be expressed in various equivalent ways
whose equivalence is not always apparent. Moreover, different formulae have
different numerical properties and it is not an easy task to choose the best
one. We restrict ourselves here to adiabatic formulae for two reasons.
Firstly, there is no nonadiabatic formulae in
literature available to us. Secondly, in the inner cavity the modes are
indeed nearly adiabatic, yielding the adiabatic coupling
coefficient to be an excellent approximation.
This is not true for the outer g-modes.
Therefore, we should treat the results obtained for them as rough estimates.
We will also discuss the effects of possible negligible coupling to
these modes.

The most detailed derivation of the adiabatic nonlinear oscillation equations
and the formula for the coupling coefficient is given by Van~Hoolst (1993,
1994a). This formula clearly shows symmetry properties of the coefficient, but
it is not good for numerical calculations in the case of high-order modes
because there are many big terms canceling each other. Numerically
much better formula of Dziembowski (1982) has a problem with treating the sharp
peak of the Brunt-V{\"a}is{\"a}l{\"a} frequency slightly below $r=0.1R$ (see
Fig.~7). The most appropriate for us seems to be the formula derived by Kumar
and Goodman (1996). Transformed to our notation it takes the following
form
\begin{equation}
H=\frac{1}{8\pi G<\rho>}\int{\mathcal L}_3{\rm d}V,
\end{equation}
where $H$ is the coupling coefficient introduced in Eqs.~(1),(2). The 
integration is done over the stellar volume, and the third-order Lagrangian is
given by
\begin{eqnarray}
\mathcal{L}_3&=&\xi^j_{0;i}\xi^i_g p'_{g;j}-
\frac{1}{2}(p'_{0;i;j}+\rho\Psi'_{0;i;j})\xi^i_g\xi^j_g-
p'_{0;i}\xi^i_g{\rm div}{\boldsymbol\xi}_g-
\frac{1}{2}p'_0({\rm div}{\boldsymbol\xi}_g)^2+\nonumber\\
&&+\frac{\Gamma(\Gamma-2)}{2}p\,{\rm div}{\boldsymbol\xi}_0
({\rm div}{\boldsymbol\xi}_g)^2-
\frac{1}{2}p_{;i}\xi^i_0({\rm div}{\boldsymbol\xi}_g)^2-
p_{;i}\xi^i_g{\rm div}{\boldsymbol\xi}_g{\rm div}{\boldsymbol\xi}_0-
\nonumber\\
&&-p_{;i;j}\left(\frac{1}{2}\xi^i_g\xi^j_g{\rm div}{\boldsymbol\xi}_0+
\xi^i_g{\rm div}{\boldsymbol\xi}_g\xi^j_0\right)-
\frac{1}{2}(p_{;i;j;k}+\rho\Psi_{;i;j;k})\xi^i_0\xi^j_g\xi^k_g.
\end{eqnarray}
In the last expression $\boldsymbol\xi_0$ is the acoustic mode eigenvector and
the $\boldsymbol\xi_g$ symbol denotes the sum of the gravity mode eigenvectors.
The semicolons denote the
covariant derivatives, upper indices denote contravariant vector components,
$$
p'_a=-{\boldsymbol\xi}_a\cdot{\boldsymbol\nabla}p-\Gamma p\,
{\rm div}{\boldsymbol\xi}_a
$$
is the Eulerian pressure perturbation of the mode $a$, $\Psi$ is the 
gravitational potential, and $\Psi'_0$ is its Eulerian perturbation of the
acoustic mode. We make use of the Cowling approximation for the g-modes,
$\Psi'_1=\Psi'_2=0$.

We introduce eigenfunction $$y_{2,k}=z_k{\mathcal C}\sigma_k^2,$$ 
where
${\mathcal C}=3(r/R)^3M/M_r$ and $M_r$ is the mass in a sphere of the radius 
$r$\footnote{Note that ${\mathcal A}/{\mathcal C}$ is the square of the
dimensionless Brunt-V{\"a}is{\"a}l{\"a} frequency ${\mathcal N}_{BV}$.}.
Additionally, for the acoustic mode we introduce eigenfunctions 
$y_{3,0},y_{4,0}$ defined by
\begin{eqnarray}
\Psi'_0&=&gry_{3,0}Y_0\nonumber\\
\frac{{\rm d}\Psi'_0}{{\rm d}r}&=&gy_{4,0}Y_0.\nonumber
\end{eqnarray}
Then we substitute eigenfunctions into Eq.~(81), keep only terms with
products of the functions of three different modes (only such terms
contribute to the resonant coupling), and after very long and
tedious algebra we obtain finally
\begin{eqnarray}
H&=&\frac{Z}{2}\int\frac{\rho r^4{\rm d}r}{\mathcal C}\bigg\{y_{1,1}y_{1,2}
\bigg[y_{1,0}\bigg(\mathcal{A}\left(4+{\mathcal C}\sigma_0^2-U-V_g\right)
+2{\mathcal C}\sigma_1\sigma_2(2-V_g)\bigg)+
\nonumber\\&&\qquad
+y_{2,0}\left(\mathcal{A}+2{\mathcal C}\sigma_1\sigma_2\right)
\left(V_g-\frac{\Lambda_0}{{\mathcal C}\sigma_0^2}\right)-
y_{3,0}V_g\left(\mathcal{A}+2{\mathcal C}\sigma_1\sigma_2\right)-
y_{4,0}{\mathcal A}\bigg]+
\nonumber\\&&
+y_{1,1}y_{2,2}\bigg[y_{1,0}\bigg(V_g\left(4+{\mathcal C}\sigma_0^2-U-V\right)
-\frac{\Lambda_0+\Lambda_2-\Lambda_1}{2}
\left(\frac{\mathcal A}{{\mathcal C}\sigma_0^2}+
2\frac{\sigma_1}{\sigma_2}\right)\bigg)+
\nonumber\\&&\qquad\qquad
+y_{2,0}\bigg(V_g\left(V-\frac{\Lambda_0}{{\mathcal C}\sigma_0^2}\right)
+\frac{\Lambda_0+\Lambda_2-\Lambda_1}{2}
\left(\frac{{\mathcal A}-1}{{\mathcal C}\sigma_0^2}+
\frac{1-{\mathcal C}\sigma_1^2}{{\mathcal C}\sigma_2^2}
\right)\bigg)-\nonumber\\
&&\qquad\qquad-y_{3,0}VV_g-y_{4,0}V_g\bigg]+\nonumber\\&&
+y_{2,1}y_{1,2}\bigg[y_{1,0}\bigg(V_g\left(4+{\mathcal C}\sigma_0^2-U-V\right)
-\frac{\Lambda_0+\Lambda_1-\Lambda_2}{2}
\left(\frac{\mathcal A}{{\mathcal C}\sigma_0^2}+
2\frac{\sigma_2}{\sigma_1}\right)\bigg)+
\nonumber\\&&\qquad\qquad
+y_{2,0}\bigg(V_g\left(V-\frac{\Lambda_0}{{\mathcal C}\sigma_0^2}\right)
+\frac{\Lambda_0+\Lambda_1-\Lambda_2}{2}
\left(\frac{{\mathcal A}-1}{{\mathcal C}\sigma_0^2}+
\frac{1-{\mathcal C}\sigma_2^2}{{\mathcal C}\sigma_1^2}
\right)\bigg)-\nonumber\\
&&\qquad\qquad-y_{3,0}VV_g-y_{4,0}V_g\bigg]+\nonumber\\&&
+y_{2,1}y_{2,2}\bigg[y_{1,0}\bigg(V_g^2(\Gamma-1)-V_g\left(
\frac{\Lambda_0+\Lambda_1-\Lambda_2}{2{\mathcal C}\sigma_1^2}+
\frac{\Lambda_0+\Lambda_2-\Lambda_1}{2{\mathcal C}\sigma_2^2}\right)
-\nonumber\\&&\qquad\qquad\qquad\quad
-\frac{\Lambda_1+\Lambda_2-\Lambda_0}{{\mathcal C}\sigma_1\sigma_2}
\bigg)+\nonumber\\&&\qquad\quad
+y_{2,0}\bigg(V_g^2(1-\Gamma)+V_g\left(
\frac{\Lambda_0+\Lambda_1-\Lambda_2}{2{\mathcal C}\sigma_1^2}+
\frac{\Lambda_0+\Lambda_2-\Lambda_1}{2{\mathcal C}\sigma_2^2}\right)
+\nonumber\\&&\qquad\qquad\qquad
+\frac{(\Lambda_1-\Lambda_2)^2-\Lambda_0^2}
{2{\mathcal C}^2\sigma_0^2\sigma_1\sigma_2}\bigg)+\nonumber\\
&&\qquad\quad+y_{3,0}\bigg(V_g^2(\Gamma-1)-V_g\left(
\frac{\Lambda_0+\Lambda_1-\Lambda_2}{2{\mathcal C}\sigma_1^2}+
\frac{\Lambda_0+\Lambda_2-\Lambda_1}{2{\mathcal C}\sigma_2^2}\right)\bigg)
\bigg]\bigg\},
\end{eqnarray}
where
$$Z=\int Y_0^*Y_1Y_2{\rm d}\Omega,\qquad U=\frac{4\pi\rho r^3}{M_r}.$$
If the acoustic mode is radial the linear Eulerian perturbation of the
gravitational potential can be obtained analytically and the function
$y_{2,0}$ has a 
different meaning than for the nonradial modes. It changes a bit the final
formula for the coupling coefficient. It turns out, however, that if we put
$y_{3,0}=0$ and $y_{4,0}=-Uy_{1,0}$, Eq.~(82) is valid in this case, too.

In the asymptotic approximation we have to substitute Eqs.~(74)--(76) into
Eq.~(82). The products of trigonometric functions of phases $\psi_{1,2}$ can
be presented as sums or differences of trigonometric functions of
$\psi_1+\psi_2$ and $\psi_1-\psi_2$,
\eg $\sin\psi_1\cos\psi_2=1/2[\sin(\psi_1+\psi_2)+\sin(\psi_1-\psi_2)]$.
We ignore terms varying with the rapidly changing phase
$\psi_1+\psi_2$ which nearly cancels out upon integration.
We see that only modes with not too large difference
of the radial orders $n$ couple effectively because only then the phase 
difference changes slowly. The final formula for the coupling coefficient is 
obtained after keeping the dominant terms in $\Lambda_{1,2}$
\begin{eqnarray}
H&=&\frac{Z{\mathcal D}_1{\mathcal D}_2}{4}\int
\frac{{\rm d}r}{r\sqrt{\kappa_1\kappa_2}}
\bigg\{\cos(\psi_2-\psi_1)\bigg[y_{1,0}\bigg(\frac{\mathcal A}{\mathcal C}
\left(4+{\mathcal C}\sigma_0^2-U-V_g\right)+\nonumber\\
&&\qquad\qquad\qquad\qquad\qquad\quad
+2\sigma_1\sigma_2(2-V_g)-\kappa_1\kappa_2\sigma_1\sigma_2
\frac{\Lambda_1+\Lambda_2-\Lambda_0}{\Lambda_1\Lambda_2}\bigg)+\nonumber\\
&&\quad+y_{2,0}\bigg(\left(\frac{\mathcal A}{\mathcal C}+
2\sigma_1\sigma_2\right)
\left(V_g-\frac{\Lambda_0}{{\mathcal C}\sigma_0^2}\right)
+\frac{\kappa_1\kappa_2\sigma_1\sigma_2}{2{\mathcal C}\sigma_0^2}\;
\frac{(\Lambda_1-\Lambda_2)^2-\Lambda_0^2}{\Lambda_1\Lambda_2}\bigg)-
\nonumber\\&&\quad-y_{3,0}V_g\left(\frac{\mathcal A}{\mathcal C}+
2\sigma_1\sigma_2\right)-y_{4,0}\frac{\mathcal A}{\mathcal C}\bigg]+
\nonumber\\&&+
\sin(\psi_2-\psi_1)\bigg[y_{1,0}\bigg(\left(2\sigma_1\sigma_2+
\frac{\sigma_2^2}{\sigma_0^2}\;\frac{\mathcal A}{\mathcal C}\right)
\frac{\kappa_2}{\Lambda_2}\;
\frac{\Lambda_0+\Lambda_2-\Lambda_1}{2}-\nonumber\\
&&\qquad\qquad\qquad\qquad\ 
-\left(2\sigma_1\sigma_2+
\frac{\sigma_1^2}{\sigma_0^2}\;\frac{\mathcal A}{\mathcal C}\right)
\frac{\kappa_1}{\Lambda_1}\;\frac{\Lambda_0+\Lambda_1-\Lambda_2}{2}\bigg)+
\nonumber\\&&\quad+y_{2,0}\bigg(\left(\sigma_1^2-\frac{1}{\mathcal C}+
\frac{\sigma_2^2}{\sigma_0^2}\;
\frac{1-{\mathcal A}}{\mathcal C}\right)
\frac{\kappa_2}{\Lambda_2}\;
\frac{\Lambda_0+\Lambda_2-\Lambda_1}{2}-
\nonumber\\&&\qquad\quad\ -\left(\sigma_2^2-\frac{1}{\mathcal C}+
\frac{\sigma_1^2}{\sigma_0^2}\;
\frac{1-{\mathcal A}}{\mathcal C}\right)
\frac{\kappa_1}{\Lambda_1}\;
\frac{\Lambda_0+\Lambda_1-\Lambda_2}{2}
\bigg)\bigg]\bigg\}.
\end{eqnarray}
To obtain the expression for the coupling coefficient, $C$, we
still need the asymptotic formula for the moments of inertia.
Using Eqs.~(73)--(76) in the general expression,
$$I_k=\int|{\boldsymbol\xi}_k|^2\rho{\rm d}V=
\int|Y_k|^2{\rm d}\Omega
\int\left(y_{1,k}^2+\Lambda_kz_k^2\right)\rho r^4{\rm d}r,
$$
we get
\begin{equation}
I_k=\frac{{\mathcal D}_k^2}{2}\int|Y_k|^2{\rm d}\Omega
\int\left(\frac{1}{\kappa_k^2}+\frac{1}{\Lambda_k}\right){\rm d}\psi_k.
\end{equation}
We adopt the normalization of spherical harmonics such that the angular
integral in Eq.~(84) is equal $4\pi$. Together with the standard 
normalization of radial eigenfunctions, $y_{1,0}(R)=1$, this gives us a 
simple interpretation of the amplitude $A_0$ (see Eq.~72). Its absolute value,
$B_0$, is then the surface averaged value of the $\delta R/R$ amplitude. Now,
looking at Eqs.~(3),(83),(84), we see that the coefficient $C$ is 
independent of ${\mathcal D}_{1,2}$. It is also almost independent of gravity
mode degrees, $l_{1,2}$, in the limit of
$l_{1,2}\gg l_0$. The dependence on $l_0$ and $m_0$ remains. The latter comes
only from the angular integral $Z$ (see Dziembowski 1982 for explicit 
expressions).

Only some sets of values of angular degrees give the nonzero coupling.
The case of $Z=0$ takes
place if $|l_2-l_1|>l_0$ or if $(-1)^{l_2-l_1}\ne(-1)^{l_0}$. Since we study
acoustic modes with $l_0\le2$ we need to consider the cases $l_2=l_1$ for
$l_0=0$ modes, $l_2=l_1+1$ for $l_0=1$ modes, and $l_2=l_1$ and $l_2=l_1+2$ 
for $l_0=2$ modes. Additionally, the azimuthal numbers $m$ have to satisfy
the condition $m_0=m_1+m_2$.

In our numerical calculations
for each acoustic mode we consider a set of inner g-mode pairs of $l_1$
in the range from $40$ to $1000$, and for each $l_1$ the range of
$\Delta n\equiv n_2-n_1$ is from $-20$ to $20$. Furthermore, for each
$(l_1,\Delta n)$ pair we consider various $(m_1,m_2)$ pairs giving nonzero
coupling coefficient. The dependence on $l_1$ is indeed
weak and is illustrated in Fig.~8 for the lowest
unstable radial mode and $l_0=2,m_0=0$ mode with $\sigma_0=3.883$. 

\begin{figure}[!htb]
\centerline{\mbox{\includegraphics[width=0.5\textwidth]{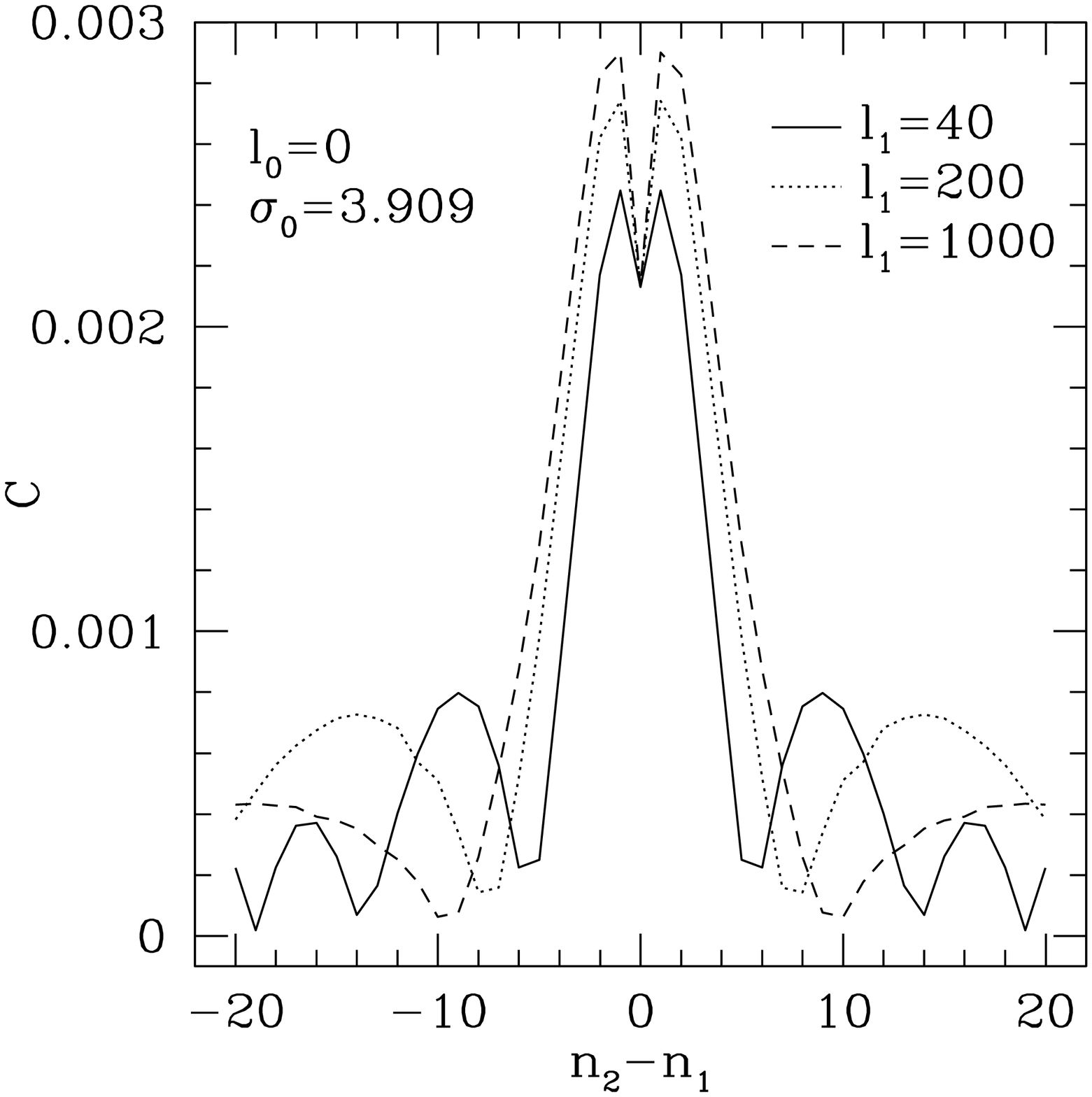}
\includegraphics[width=0.5\textwidth]{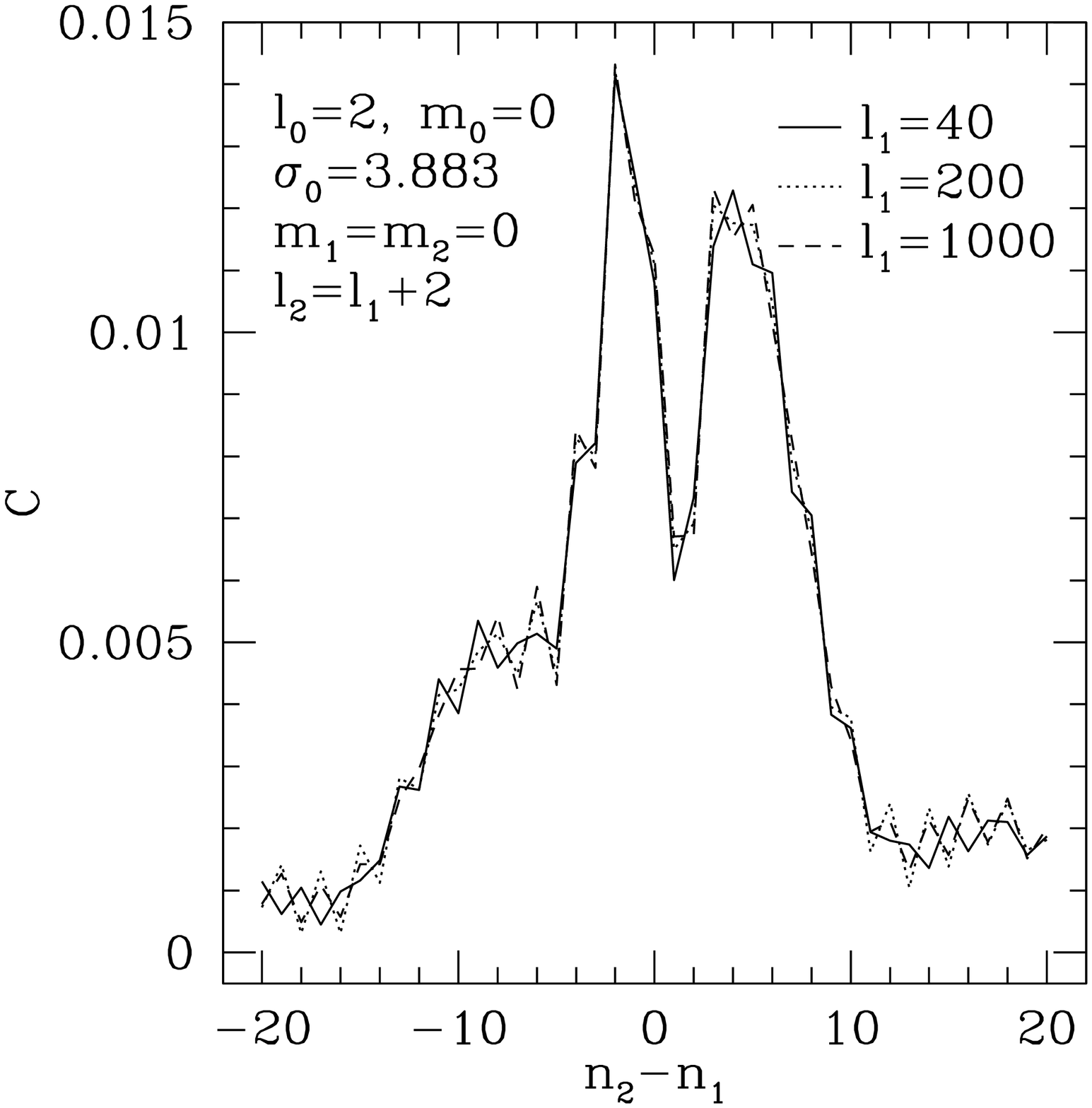}}}
\FigCap{The coupling coefficient $C$ for the inner g-mode pairs and the
lowest unstable radial mode (left panel) and $l_0=2,\ m_0=0,\ \sigma_0=3.883$
mode (right panel). In the latter case $l_2=l_1+2$ and the
g-mode azimuthal degrees are $m_1=m_2=0$ which give maximum value of $|Z|$ for
these $l$-values.}
\end{figure}

For the radial mode
(left panel) most of the coupling comes from the region of a broad maximum of
the Brunt-V{\"a}is{\"a}l{\"a} frequency, $0.1\lesssim r/R\lesssim0.5$ and
if the modes $1,2$ have different frequencies, their propagation zones have
different sizes. These differences decrease with increasing $l_1$ since the
spectra densities increase. This is the cause of a bit different 
$\Delta n$-dependence of the coupling coefficients for different $l_1$. 

The two lowest $l_0=2$ modes (one of them is in the right panel) are in
fact mixed modes with quite large values of eigenfunctions in the deepest
part of the radiative region. Then, most of the coupling
comes from the region $r/R<0.1$, and the coupling coefficient is almost
independent of the upper limit of the propagation zone. Therefore, the 
coupling coefficient dependence on $\Delta n$ is nearly the same for 
different $l_1$ numbers.

We can also see in the plot that if $l_1=l_2$ (left panel) the
coupling coefficient is symmetric about $n_2-n_1=0$. It is expected because
the change of the sign of $\Delta n$ means just interchange of the two 
g-modes and the coupling coefficient is symmetric with respect to it. 
Obviously, this is not true for $l_1\ne l_2$ (right panel).

The coupling coefficients have maxima near $n_1=n_2$ for each set of
$l_0,\sigma_0,l_2-l_1$. For the radial modes these maxima are between 
$2.7\times10^{-3}$ at $\sigma_0=3.909$ and $1.3\times10^{-3}$ at 
$\sigma_0=5.859$. In the case of $l_0=1,m_0=0$ the maxima are 
between $1.1\times10^{-3}$ at $\sigma_0=4.125$ and $0.4\times10^{-3}$ at
$\sigma_0=6.162$.
For $l_0=1,|m_1|=1$ the coupling coefficients are about 30\% larger due to
larger values of $Z$.
For $l_0=1$, the values of coupling coefficients are a few times
lower than in the case of radial modes due to lower p-mode amplitudes in
the g-mode cavity. See also Fig.~9.

As has already been mentioned, in the case of $l_0=2$ we have to consider 
$l_2=l_1$ together with $l_2=l_1+2$. Typically, the $l_2=l_1+2$ case is more
important because the coupling coefficients are higher.
The coupling coefficient maxima at $m_0=0$ and $m_0=1$
are between about $1.4\times10^{-2}$ at $\sigma=3.883$ and $1\times10^{-3}$
at $\sigma_0=5.793$ while for $m_0=2$ the maxima
are higher by a factor $\sim1.5$. The coupling coefficients
for the two lowest $l_0=2$ modes are significantly larger than the
coefficients for the remaining modes because these two modes
have mixed character and high values of eigenfunctions in the region around
$r/R=0.1$.

\begin{figure}[!htb]
\centerline{\mbox{\includegraphics[angle=270,width=1\textwidth]{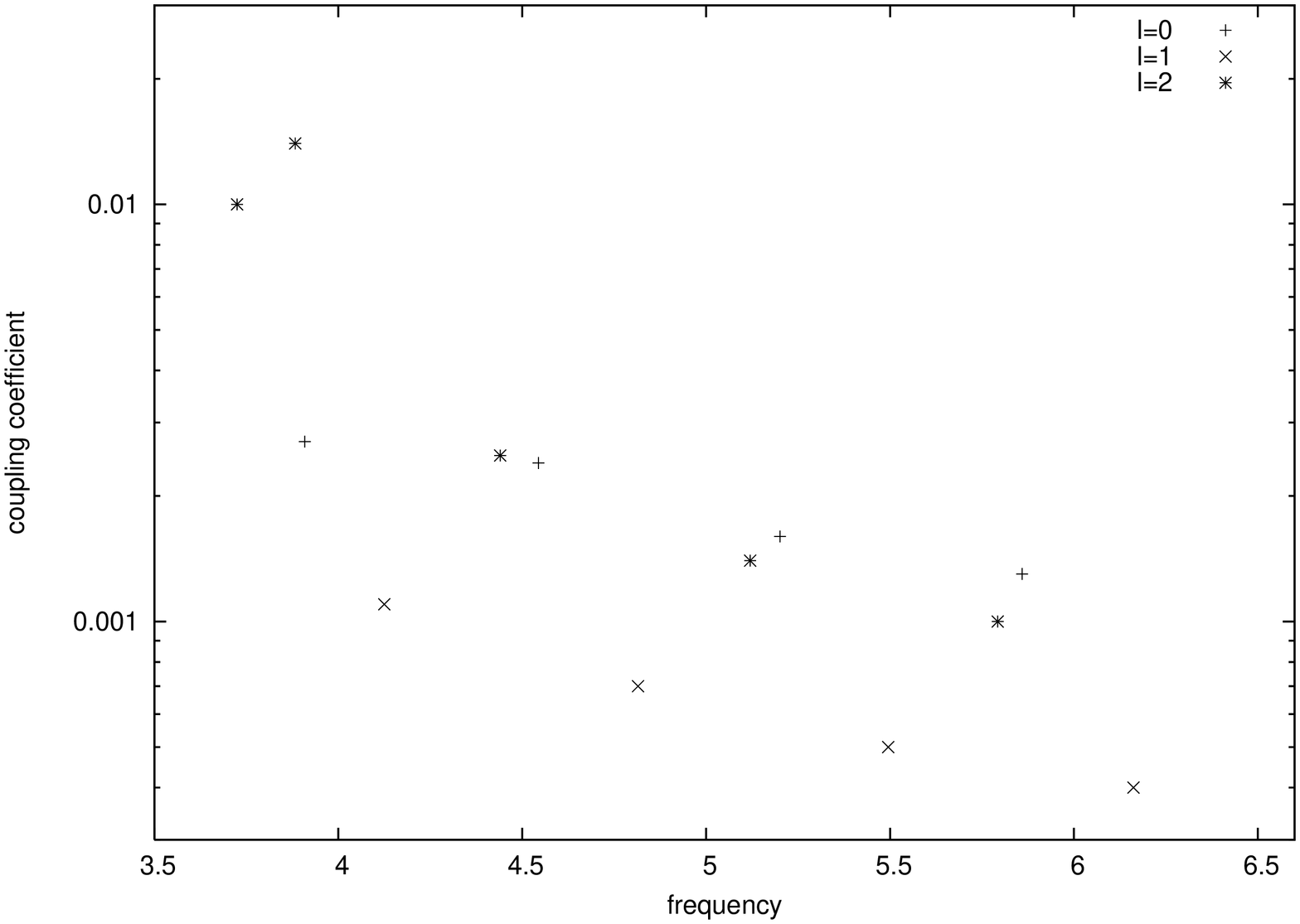}
}}
\FigCap{The coupling coefficients maxima for the acoustic modes with $m=0$.
For $l=1,m=1$ the coupling coefficients are about $30\%$ higher than those
for $l=1,m=0$. For $l=2,m=1$ the coupling coefficients are similar to those
at $l=2,m=0$. For $l=2,m=2$ the coupling coefficients are higher about the
factor $\sim1.5$ than those for $l=2,m=0,1$.}
\end{figure}

In the case of the outer gravity modes we have to use Eq.~(82) for $H$
and the general expression for
the moment of inertia. The angular degree dependence is the same as
for the inner g-modes. It turns out that only modes with frequencies 
$\sigma_{1,2}\approx\sigma_0/2$ and $n_{1,2}=1$ are important. The
estimated values of coupling coefficients for the modes in Table~1 range from 
about $7$ at $\sigma_0\lesssim4$ to about $40$ at $\sigma_0\gtrsim5.5$.
We should stress that these values should be treated as rough 
estimates, thus we do not give precise values.

\Subsection{Detuning Parameters and Critical Amplitudes}

Apart of damping rates and coupling coefficients we need values of
the detuning parameters, $\Delta\sigma$. We notice first that in the absence
of rotation the mode frequencies are independent of $m$. This means that the
detuning parameters are independent of $m$, too. It is important because
each set $l_0,m_0,l_1,l_2$ characterizes whole multiplet of g-mode pairs.
The pairs within the multiplet differ by $m$ numbers of g-modes, but each
pair has the same value of $m_1+m_2$. The pairs in the multiplet have the
same values of $\Delta\sigma$, so, in view of what was said previously about
synchronization, we may treat whole multiplet as a single pair. If $l_0\ne0$
the pairs within the multiplet have different coupling coefficients, so
we take into account only pair with the maximum value of the coefficient.

The inner g-modes with given angular degree, $l$, and different radial 
orders, $n$, form dense frequency spectra. In such a case the determination
of the detuning parameter would require an unrealisticly precise knowledge
of the eigenfrequencies. Instead, we adopt statistical approach. For each
g-modes'
angular degree, $l$, we make a random choice of position of the
frequency spectrum relative to the frequency of the acoustic mode. Then we
compute detuning parameters for various g-mode pairs. 
The relatively strong coupling takes place only for $n_1\approx n_2$ so we
may neglect $|n_1-n_2|\gtrsim10$ (see Fig.8). This means also that the relevant
pairs have $\sigma_1\approx\sigma_2$. Moreover, we explain below that only
one pair for each $(l_1,l_2)$ is important.

At given $l$ the frequencies of g-modes with consecutive $n$ may be treated
as equidistant in a relevant range of radial orders. Since $l_1\approx l_2$
and $n_1\approx n_2$, we have from Eq.~(78)
$\delta\sigma_{l_1,n_1}\approx\delta\sigma_{l_2,n_2}$, and
\begin{eqnarray}
\Delta\sigma_{l_1,n_1+1,l_2,n_2-1}&=&\sigma_0-(\sigma_{l_1,n_1+1}+
\sigma_{l_2,n_2-1})\approx\nonumber\\
&\approx&\sigma_0-(\sigma_{l_1,n_1}-\delta\sigma_{l_1,n_1}+\sigma_{l_2,n_2}+
\delta\sigma_{l_2,n_2})\approx\nonumber\\
&\approx&\Delta\sigma_{n_1,n_2}.
\end{eqnarray}
This implies that all the relevant pairs with even $n_1-n_2$ have close
detuning parameters and may be treated as synchronized. As was explained in
Section~4 such pairs may be treated as one effective pair with the coupling
coefficient equal to the maximum value among even $n_1-n_2$ pairs. Obviously,
the same is true for odd $n_1-n_2$ pairs. Finally, because
$$|\min(\Delta\sigma)_{{\rm even\ }n_1-n_2}-
\min(\Delta\sigma)_{{\rm odd\ }n_1-n_2}|\approx
\delta\sigma_{l_{1,2},n_{1,2}}$$
is much higher than $\gamma_0$ for the relevant g-mode pairs, \ie those with
$l$ of the order of a few hundreds (Eq.~78),
we may take into account only this pair which has smaller $|\Delta\sigma|$.
The other one has too high detuning parameter and is not excited.

Even the lowest value of $|\Delta\sigma|$ is often much higher than
$\gamma_0$, and then we totally
neglect g-mode pairs at a given $(l_1,l_2)$. What remains and is taken to
further computations are single pairs at some values of $l_1$.

It turns out that the most important pairs are in the range
$100\lesssim l_{1,2}\lesssim300$.
The lower-$l$ pairs are negligible for two reasons. First, their spectrum is
sparse and few of them have $\Delta\sigma\sim\gamma_0$. Second,
they have very small damping rates, which implies they hardly contribute
to the overall damping (Eq.~48). On the other hand the pairs with high
$l_{1,2}$ are strongly damped and turn out to be inactive in
time-dependent solutions.

When the rotation is taken into account, the number of independent
pairs increases due to splitting of g-mode frequencies. This effect will be 
studied in more details in Subsection~5.5.2.

In the case of the outer g-modes the damping rates are much higher than the
acoustic mode driving rates. This implies that only one pair suffices to
halt the acoustic mode growth and the system reaches
the limit cycle solution, similar to that studied in Subsection~3.1. In this
case the critical amplitude (Eq.~7) is a
good estimate of the average acoustic mode amplitude. This critical amplitude
is insensitive to the detuning parameter because it is much smaller than
the damping rate and may be neglected in Eq.~(7). This also implies that
taking into account rotation does not change the expected acoustic mode
amplitude.

The outer g-mode pairs have to be taken into account if the average
acoustic mode amplitude determined by the interaction with only the inner
g-mode pairs is higher than the critical amplitude for the excitation
of the outer g-mode pair. In particular, such a situation takes place
for the acoustic modes with relatively high growth rates, because then
many inner g-mode pairs are necessary to halt the driving. The detuning
parameters of those pairs are also relatively high. Even though there is no
simple dependence of resulting average acoustic mode amplitude on detuning
parameters of active inner g-mode pairs, the qualitative dependence is
similar as in the single-pair case, \ie the
higher detuning parameters are, the higher the amplitude is. In our model,
higher frequency acoustic modes have high growth rates and we expect
high average amplitudes determined by the interaction with the inner
g-modes. Then, the outer g-mode pairs may dominate the interaction.

\Subsection{Time Evolution}
 
\subsubsection{The Case of Slow Rotation}

In the case of sufficiently slow rotation, the splitting of the g-modes is
smaller than their damping rates and the multiplets may be treated as single
modes, as was explained in Subsection~3.3. Here we study a few
examples of such systems. In all cases we set $m_0=0$. Other values of $m_0$
give slightly different values of coupling coefficients, which results only
in slight change of the amplitudes. Indeed, if we increase the coupling
coefficients in Eqs.~(30)--(32) by some factor and decrease the amplitudes
by the same factor, the equations will remain unchanged.
This means the amplitude values scale with
coupling coefficients without change of the time dependence which is
governed by the p-mode growth rate.

At first, we consider the $l_0=2,\sigma_0=3.883$ mode together with an
ensemble of several hundred inner g-mode pairs in the $l_1$-range between $40$
and $1000$. The beginning of the time evolution of the
acoustic mode amplitude is shown in Fig.~10.
The initial value of the amplitude is set to be equal to the lowest critical
amplitude. The initial g-mode energies were randomly distributed between zero
and the acoustic mode initial energy. At the very beginning the acoustic
mode amplitude drops as the result of the interaction with hundreds of
artificially excited
g-modes. Then, the modes become temporarily uncoupled and g-mode amplitudes
fall down, while the acoustic mode amplitude starts to grow exponentially.
Next, several g-mode pairs become excited due to the interaction with the
acoustic mode and a statistical equilibrium is reached after about
$5\times10^4$ days.
The average amplitude in this state is about $0.001$ and average modulation
period is about $3000$ days. It is of the order of $1/\gamma_0\approx5600$
days. Only inner g-modes actively interact with the acoustic mode.

\begin{figure}[!htb]
\centerline{\mbox{\includegraphics[width=0.7\textwidth]{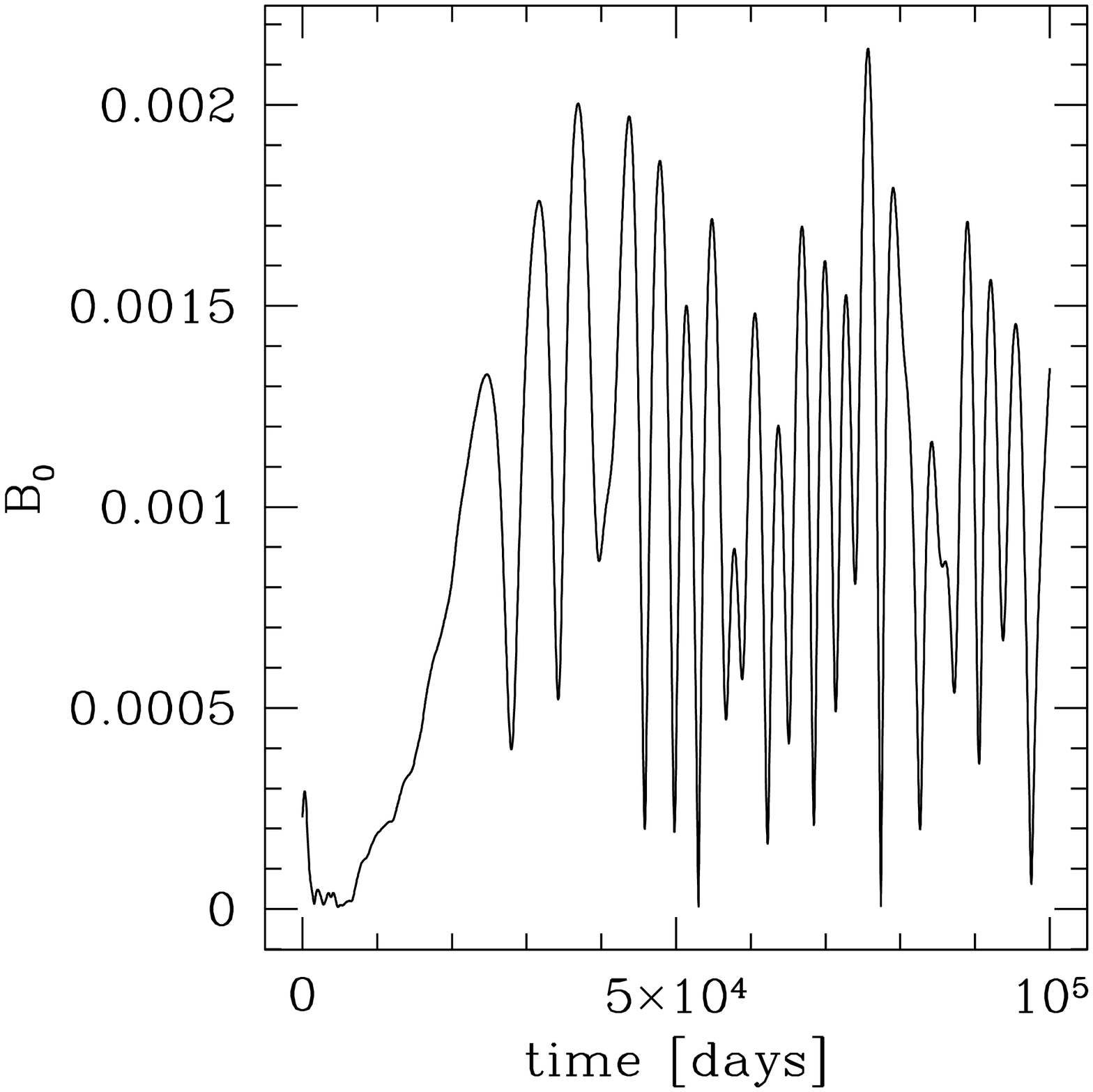}
}}
\FigCap{The time evolution of the $l_0=2,\sigma_0=3.883$ acoustic mode 
amplitude.}
\end{figure}

The $l_0=2,\sigma_0<4$ modes are not typical unstable modes in our model.
They are, in fact, mixed modes with significant amount of energy concentrated
in the internal g-mode cavity. This causes strong coupling to the inner
g-modes and weak linear driving, the latter being the result of a large mode
inertia. Therefore, we performed the same time integration
for the radial $\sigma_0=3.909$ mode. Although the outer g-modes should be
taken into account in this case, we first consider
only inner g-modes in order to compare the results with the previous
case when the growth rate was significantly smaller.

The time evolution of the radial mode amplitude in the statistical
equilibrium without
outer g-modes is presented in Fig.~11. We see that the average amplitude is 
about $0.01$ and the typical modulation period is about $1000$ days. This
timescale is again of the order of $1/\gamma_0\approx1600$ days. The
amplitude is about $10$ times higher than in the case in Fig.~10, mainly
due to coupling coefficients which are smaller by a factor of $5$.
The remaining factor of $2$ turns out to be the result of dependence of the
average amplitude on the growth rate. This dependence is expected because
$\gamma_0$ determines $\Delta\sigma$ range of active g-mode pairs and, as in
the case of a single pair, there is some dependence of the average amplitude
on detuning parameters.
\begin{figure}[!htb]
\centerline{\mbox{\includegraphics[width=0.7\textwidth]{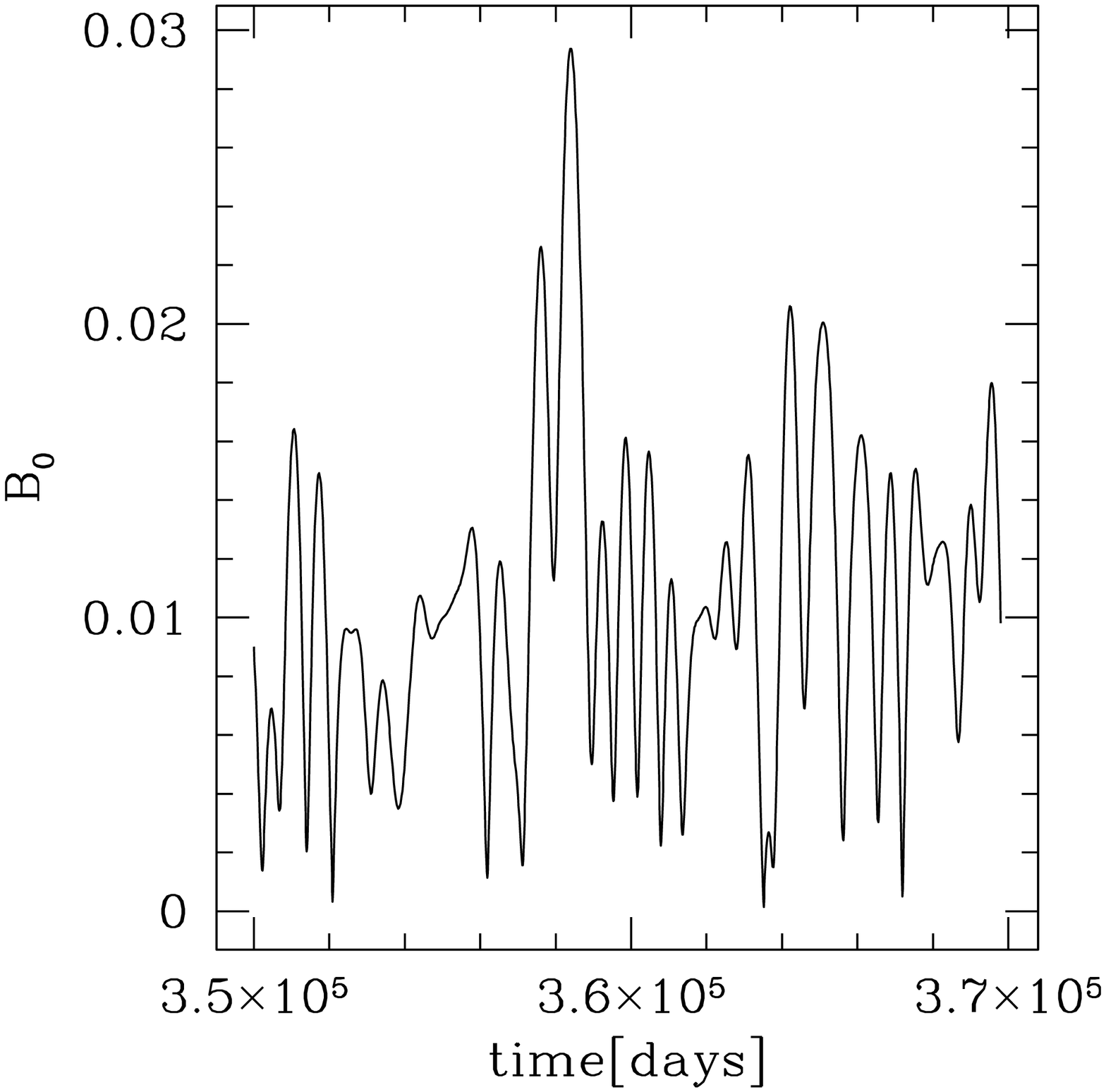}
}}
\FigCap{The time evolution of the $l_0=0,\sigma_0=3.909$ acoustic mode 
amplitude without outer g-modes.}
\end{figure}

The time evolution of the same radial mode with the outer
g-modes taken into account is presented in Fig.~12.
There is a qualitative difference in amplitude behavior between
this case and the previous one. This difference is caused by the presence of
outer g-modes which are strongly damped and strongly coupled to the radial
mode. The sudden jumps of amplitude, lasting less than 10 days,
coincide with very short time intervals during which the outer g-mode is
excited. 
\begin{figure}[!htb]
\centerline{\mbox{\includegraphics[width=0.7\textwidth]{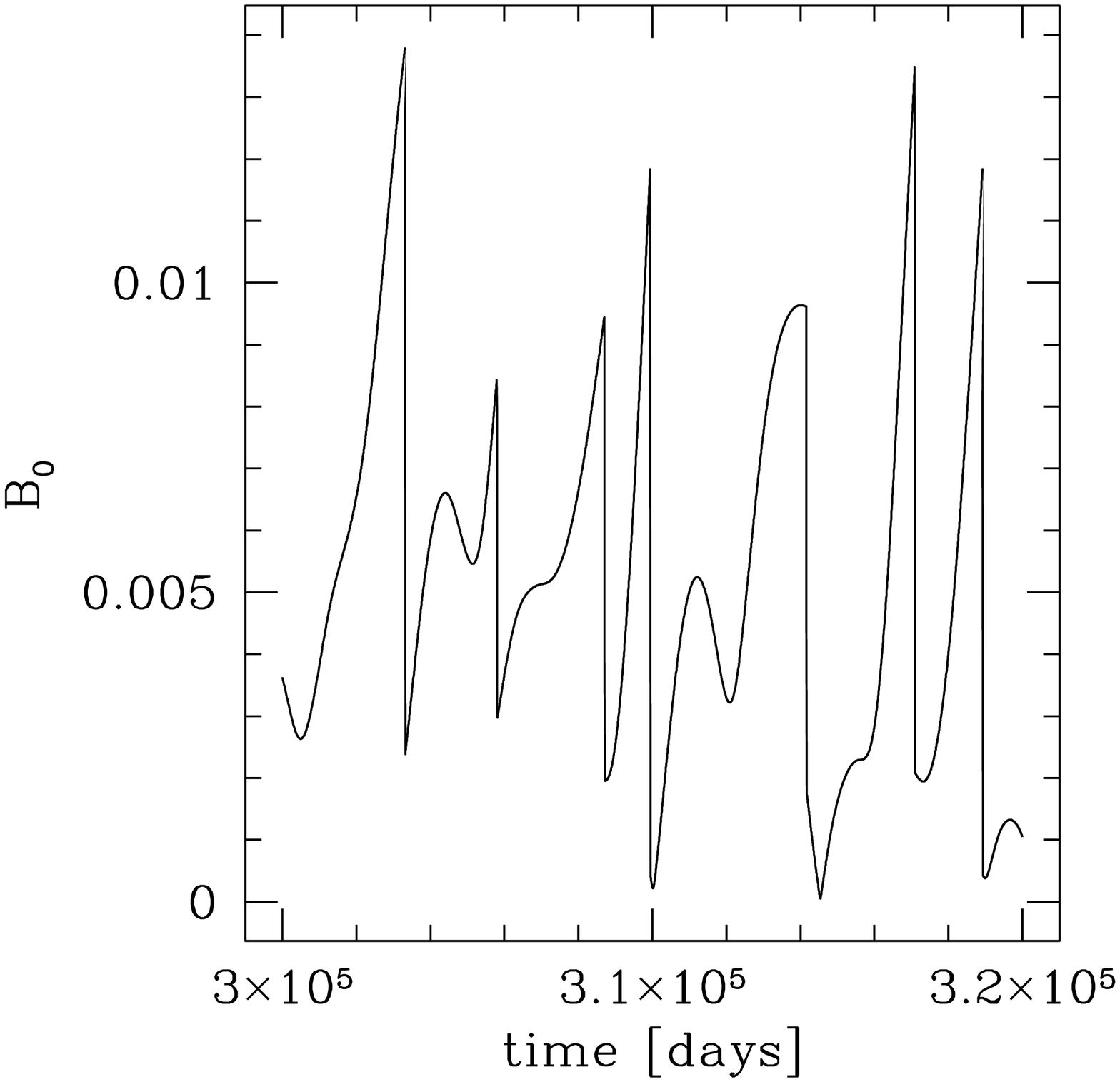}
}}
\FigCap{The time evolution of the $l_0=0,\sigma_0=3.909$ acoustic mode 
amplitude with outer g-modes.
}
\end{figure}

Different behavior of the modes in this case than in the previous ones
is caused by orders-of-magnitude larger ratio
$|\gamma_{1,2}|/\gamma_0$ of the dominant g-modes. Some of the inner g-mode
pairs also remain excited, which is manifested by smooth
amplitude changes, \eg around $3.12\times10^5$ days. However, the inner
g-modes play a secondary role in this case. In spite of different behavior
of the amplitude, the modulation timescale is similar to that in the case
without
inner g-modes. This is the manifestation of the fact that also in this case
the timescale is governed by the acoustic mode growth rate.

This case resembles
the case of a single g-mode pair, because only one pair is sufficient to
stabilize the acoustic mode. The average amplitude of the latter is of the
order of the critical amplitude to the excitation of the outer g-mode, equal
approximately $0.005$. This value is smaller than the average amplitude in the
previous example, \ie determined by inner g-modes and equal $\sim0.01$.
Therefore, the interaction with the outer g-mode dominates.

The last example is the case of the $l_0=1,\sigma_0=5.495$ acoustic
mode which is one of the most strongly unstable modes and for which the
coupling to the inner g-modes is one of the weakest. The time evolution of
this mode is shown in Fig.~13. Only an outer g-mode pair remain
effectively interacting after several hundred days. The system converges
to the periodic limit cycle with this pair. Very short modulation timescale,
about $100$ days, is caused by very high growth rate, $1/\gamma_0\approx50$
days.

\begin{figure}[!htb]
\centerline{\mbox{\includegraphics[width=0.7\textwidth]{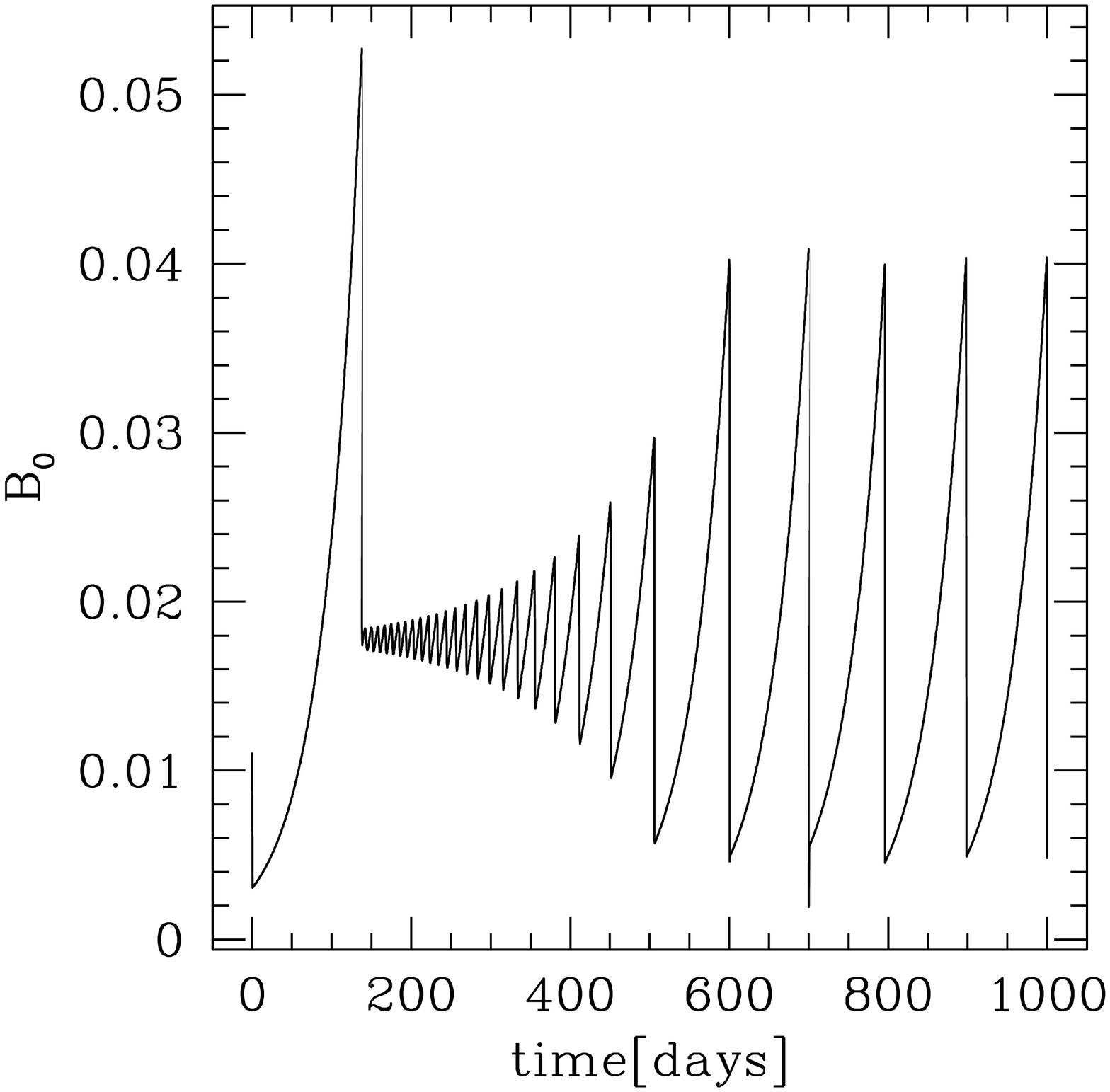}}}
\FigCap{The time evolution of the $l_0=1,\sigma_0=5.495$ acoustic mode 
amplitude.}
\end{figure}

The high value of $\gamma_0$ demands many weakly damped inner g-mode pairs
to stabilize the system. Moreover, these modes are very weakly coupled to the
acoustic mode and, consequently, the resulting average amplitude of the latter
would be much higher than $\sim0.02$, \ie the critical amplitude for the
excitation of the outer g-mode pair. Therefore, the outer g-mode pair strongly
dominates the interaction and we cannot even notice the presence of the
inner g-mode pairs.

Other unstable acoustic modes have average amplitudes similar to those in
the last two cases, \ie of the order of $0.01\div 0.02$, and the interaction
is dominated by the outer g-modes.

\subsubsection{The Case of Moderate and Fast Rotation}

Since the modes of close frequencies effectively act as independent only if
the $\Delta\sigma$-separation is sufficiently large, rotational splitting
may significantly change the picture only if rotation is fast enough.
In this section we will try to determine the critical rotation rate at which
modes can be treated as independent. In Subsection~3.3 we showed that the
separation must exceed $\gamma$ for pairs to be regarded independent.

Now, let us consider $\Delta\sigma$ separation between members of g-mode
multiplets that may be coupled to low-degree acoustic mode.
Very good approximation for the high-$l$ inner g-mode's eigenfrequency in the
corotating frame, up to quadratic terms in rotation is (Dziembowski \etal 1988)
\begin{equation}
\sigma_{\rm rot}\approx\sigma_{\rm nrot}+\frac{m\Omega_{\rm rot}}{l^2}-
\frac{m^2\Omega_{\rm rot}^2}{l^2\sigma_{\rm nrot}},
\end{equation}
where $\Omega_{\rm rot}$ is dimensionless rotational frequency of a star.
The g-modes in each pair have angular degrees
$l_1\approx l_2$ and $m_1\approx-m_2$ so we neglect the difference in linear
terms in rotation in the formula for the detuning parameter
\begin{equation}
\Delta\sigma_{\rm rot}\approx\Delta\sigma_{\rm nrot}+
\frac{4}{\sigma_0}\left(\frac{m_1\Omega_{\rm rot}}{l_1}\right)^2,
\end{equation}
where we used $\sigma_{\rm nrot}\approx\sigma_0/2$.
The rotational frequency change of the acoustic mode is unimportant here,
since it only shifts all the detuning parameters by the same value, without
changing their statistical properties. The rotation begins to be important
when
\begin{equation}
\Omega_{\rm rot}^2>\frac{|\gamma_{l_1}|\sigma_0}{4}.
\end{equation}
In the numerical experiments from the previous subsection with only inner
g-modes taken into account, the typical $l$ of
active pairs was $200$. Thus, from Eq.~(88) we find that the rotation
becomes important at $\Omega_{\rm rot}\gtrsim0.001$. Assuming uniform rotation
of the star, this value in our model corresponds to equatorial velocity
below $1$ km/s.

The linear terms in rotation are higher than the quadratic ones
for very small rotation rate and $l_1\ne l_2$. However, at $l_0=2$ and
$l_1\sim10^2$ we estimated that the
linear $\Delta\sigma$-splitting exceeds the quadratic one at rotation
rate smaller than $0.4\times10^{-3}$. At such low values the splitting is
smaller than $|\gamma_l|$ so the rotation is negligible. When the rotation
becomes significant, only quadratic terms play role.

In order to assess terminal amplitude of acoustic mode in the presence of
rotation we performed a number of numerical experiments.
To reduce the number of g-mode pairs we divide each multiplet into
groups of pairs such that detuning parameters in each group differ less than
the value of the damping rate,
\begin{equation}
|\Delta\sigma_{l,m}-\Delta\sigma_{l,m'}|<|\gamma_l|.
\end{equation}
Each group of pairs is treated as a single pair with mean detuning parameter
and coupling coefficient equal the maximum one within the
group. We ignored the influence of rotation on the coupling coefficients.
The detuning parameters of different groups differed more than the
damping rate of the multiplet. In particular, in the absence of rotation
whole multiplet formed one group of pairs and was thus replaced by a single
pair.

An example of the time evolution of the $l_0=0,\sigma_0=3.909$ mode amplitude
with the rotation frequency $\Omega_{\rm rot}=.01$ is presented in Fig.~14.
We see that the average amplitude is about $0.0025$. It is about factor
of $\sim4$ lower than in the case without rotation (see Fig.~11).
Moreover, the characteristic timescale is somewhat longer than in the
case without rotation. 
This is the result of much denser $\Delta\sigma$ spectrum of active pairs.
The dense spectrum implies that there are many independent low-$l$ pairs
in the vicinity of exact resonance and there is no need to excite pairs with
higher $\Delta\sigma$. Consequently, the modulation timescales are shorter,
but still they are of the order of $1/\gamma_0$.
\begin{figure}[!p]
\centerline{\includegraphics[width=0.7\textwidth]{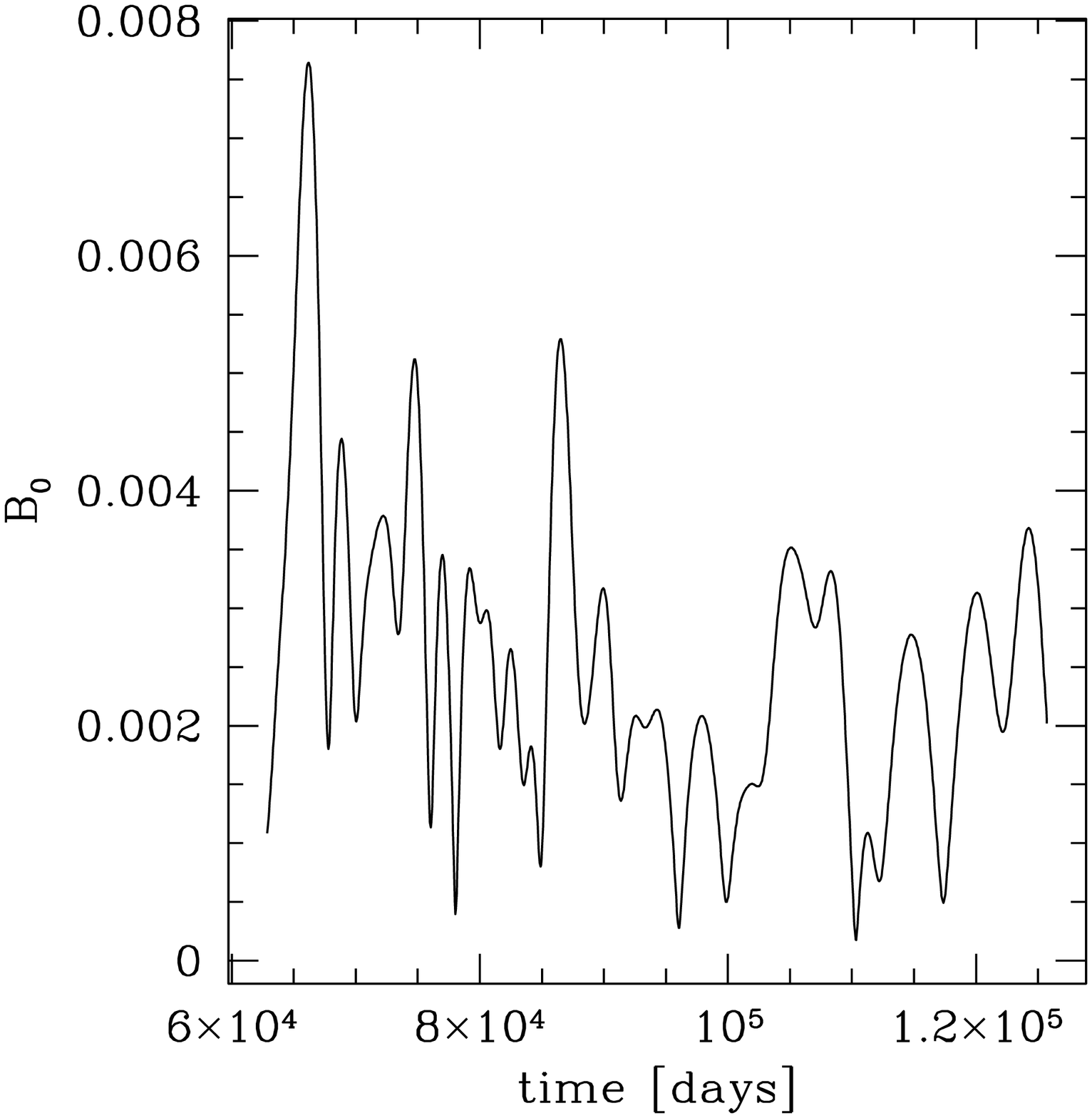}
}
\FigCap{The time evolution of the $\sigma_0=3.909$ radial mode interacting
with the g-mode pairs in the presence of rotation, $\Omega_{\rm rot}=0.01$.
}
\end{figure}
\begin{figure}[!p]
\centerline{\includegraphics[width=0.7\textwidth]{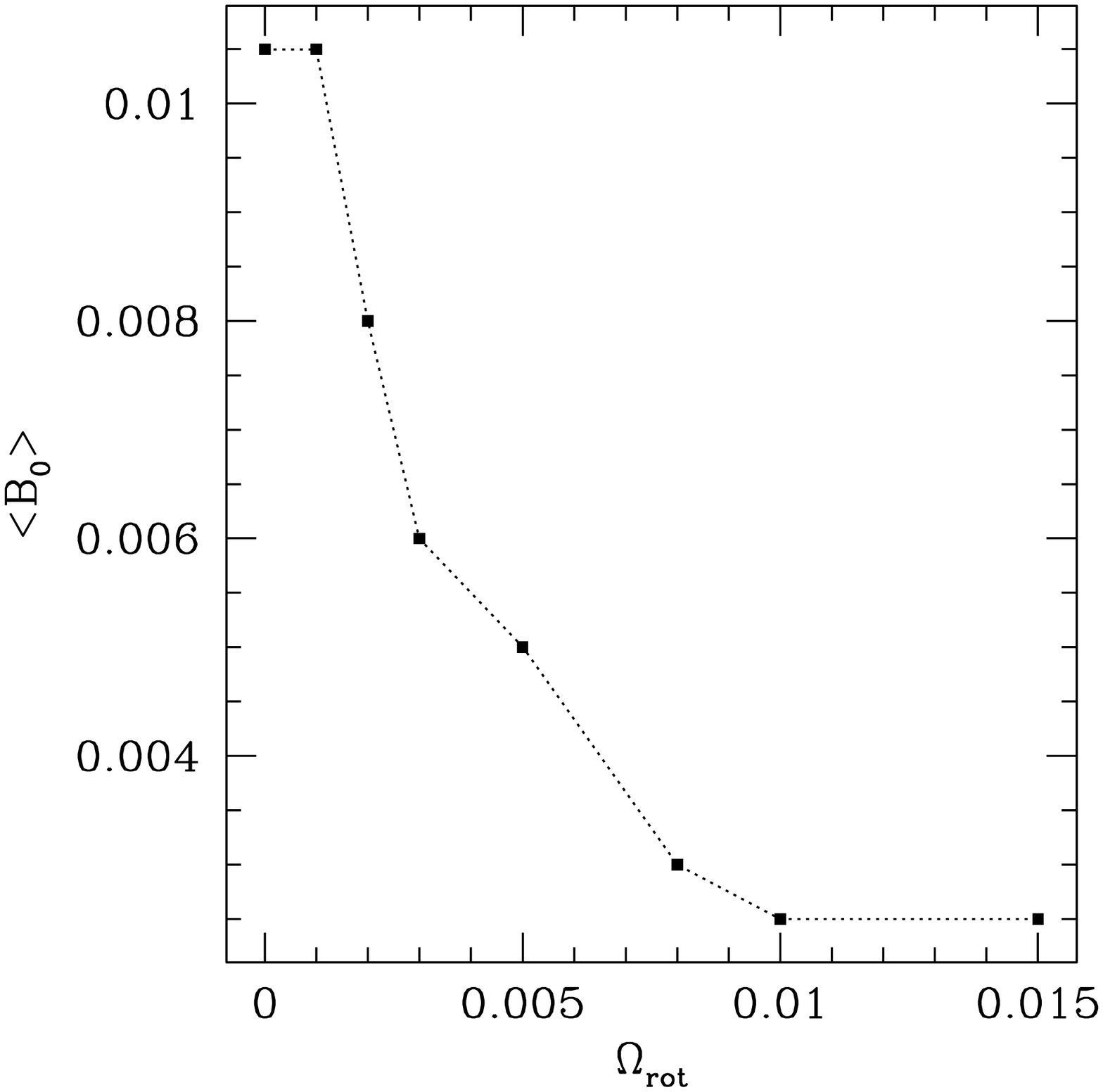}
}
\FigCap{Average amplitude of the $\sigma_0=3.909$ radial mode interacting with
the inner g-modes versus dimensionless stellar rotation rate.}
\end{figure}

The transition from no rotation to fast rotation is shown in Fig.~15. At very
slow rotation its influence on the average amplitude is negligible.
The rotation becomes significant at $\Omega_{\rm rot}\approx0.002$
which is of the order of the simple estimate obtained above.
Further increase of the rotation rate causes that the average acoustic mode
amplitude decreases. Above $\Omega_{\rm rot}\approx0.01$ the effect of
rotation saturates. We may treat the rotation rate at which the average
amplitude is reduced by the factor of $2$ as the limit between slow and
fast rotation. This limit in our case is $\Omega_{\rm rot}\approx0.003$.
Assuming uniform rotation this value corresponds to equatorial velocity
about $2$ km/s.

If we take into account outer g-modes it turns out that they dominate at slow
rotation (see previous subsection). At fast rotation inner g-modes are able to
limit amplitude growth
of the radial mode at values below critical ones for outer g-modes which thus
remain unexcited.

The role of rotation is important for the p-modes whose amplitudes are
determined by the interactions with the inner g-modes. In our model they are
weakly unstable $l_0=2,\sigma_0<4$ modes. Their amplitudes are reduced by
the factor of a few due to fast rotation.

On the other hand, for high frequency p-modes ($\sigma_0>4$) even fast
rotation cannot change the situation, because even at the fastest rate the
outer g-modes must be excited and they determine the p-mode amplitudes.
As was explained in Subsection~5.4, the outer g-modes are insensitive to
rotation.
Thus the values of the amplitudes of the $\sigma_0>4$ p-modes are the same as
in Subsection~5.5.1, \ie of the order of $0.01\div 0.02$.

\Subsection{Luminosity Amplitudes}

In order to obtain the expected luminosity amplitudes, we have to multiply
radius variations amplitudes by a nonadiabatic coefficient $f$. As was
mentioned in Subsection~5.1, the absolute value of this coefficient for all
considered acoustic
modes is about $30$. The effects of flux averaging over the stellar disc play
role mainly for $l=2$ modes, and they reduce effective $|f|$ to the
value of about $20$. Moreover, for the modes whose radius amplitudes are
determined by the interaction with the inner g-modes, we adopt the values
obtained at fast rotation.

In the case of radial acoustic modes the evaluated bolometric luminosity
amplitudes range from about $0.07$ mag at $\sigma=3.909$ to more than
$0.5$ mag at $\sigma=5.859$. Only the $\sigma=3.909$ radial mode has
amplitude determined by the interaction with the inner g-modes and
the value of the amplitude is calculated with the rotation taken into account.

All the $l=1$ acoustic modes have amplitudes determined by the outer
g-modes. The values of the amplitudes range from $\sim0.15$ mag at
$\sigma=4.125$ to more than $0.5$ mag at $\sigma>5$.

The $l=2$, acoustic modes with $\sigma<4$ have relatively small expected
luminosity amplitudes, about $5$ mmag. This is the combined result of their
low growth rates, strong coupling to the inner g-modes, fast rotation and 
disc-averaging effect.
However, at $\sigma>4$ the expected amplitudes are again high, of the order
of a few tenths of a magnitude.

Comparing obtained luminosity amplitudes with those observed in XX~Pyx or
other $\delta$-Scuti stars, we see that only low-frequency $l=2$ modes
in our model have expected amplitudes below $10$ mmag. Higher frequency
acoustic modes have amplitudes equal a large fraction of a magnitude.
These values are typical for HADS stars.
Neglecting outer g-modes results in even higher amplitudes.
This suggests that nonresonant saturation
effects play an important role in limiting the amplitudes.
However, at present stage
we cannot exclude the possibility
that our treatment of the outer g-modes is too crude and
more realistic modeling of those modes would give better consistency with
observations.

\Subsection{Effects of Saturation of The Linear Driving}

The saturation of the instability mechanism is manifested by the nonlinear
dependence of the growth rates on mode amplitudes.
The nonlinear growth rates may be written in the form
(see, \eg Buchler and Goupil 1984)
\begin{equation}
\gamma_k^{\rm nlin}=\gamma_k^{\rm lin}
\left(1+\sum_j S_{k,j} B_j^2\right),
\end{equation}
where $S_{k,j}<0$ are saturation coefficients and the summation is made over
all modes excited in the star.

The simplest way to consider the saturation mechanism is to reduce
the values of the growth rates.
Then, as implied by Eq.~(48), 
the number of g-modes necessary to achieve the statistical
equilibrium is reduced, too.
The lower driving rate, as we explained in Section~4, means also that
the $\Delta\sigma$ range of active pairs is smaller.
Consequently, the expected amplitude of the acoustic mode becomes lower and
the modulation timescale is longer.

The effect of the growth rate saturation is relevant only in the
interaction with the inner g-modes. For the outer g-modes there is no
reduction of the amplitude because the damping rate is always much higher than
the growth rate.

The cases of very small growth rates were studied by Dziembowski and
Kr\'olikowska (1985) and Dziembowski \etal (1988). The amplitude of the
acoustic mode is then given by the lowest critical amplitude (Eq.~7) for
the whole ensemble of the resonant g-mode pairs. It is so, because the
pair corresponding to this minimum is able to halt the acoustic mode growth.
The system reaches the stable equilibrium, described in Section~2.1.2, or a
periodic limit cycle such as described in Section~3.1. The expected amplitude
of the acoustic mode in both situations is close to the critical one.

For the XX~Pyx the simple assessment based on Eq.~7 with the rotation and
strong saturation taken into account and the respective $f$ coefficient leads
to amplitude ranging from
about $1$ mmag at $l=2,\sigma<4$ modes up to about $0.1$ mag at $l=1,\sigma>5$
modes. The amplitudes of the high frequency acoustic modes are still much
larger than observed in XX~Pyx. The fact that we find amplitudes considerably
higher than Dziembowski and Kr\'olikowska (1985) did, is a consequence of
evolved $\delta$~Scuti star model while they used a ZAMS star model. In our
case the inner g-modes are more concentrated at the peak of the
Brunt-V\"ais\"al\"a frequency below $r=0.1R$ and consequently they are much
more weakly coupled to the p-mode. The interaction with the outer g-modes
is also weaker in our case because of higher damping rates.

The unavoidable conclusion is that the dominant effect responsible for the
amplitude limitation in $\delta$~Scuti stars must be nonlinear saturation.
In such a situation our basic approximation that acoustic modes remain
uncoupled is not allowed. The coupling between various acoustic modes occurs
through the collective saturation of the driving mechanism, as described by
Eq.~(90). The decrease of amplitudes of certain modes gives the chance for
other modes to be excited. The net effect depends on the linear growth rates
and the saturation coefficients. In Fig.~6. we see that the linear growth
rates initially grow rapidly with frequency. However, the saturation
coefficients are also growing with frequencies as both observations and
hydrodynamical simulations show. The high frequency modes are both easy to
excite and easy to saturate. It requires numerical calculations to predict
the outcome of the collective saturation. The only thing that can be said at
this stage is that the amplitude changes should be faster for high frequency
modes.

\Section{Conclusions}

We studied the resonant coupling between unstable acoustic modes and pairs of
stable g-modes. We reminded the general analysis of the interaction given
by Dziembowski (1982) and Dziembowski and Kr{\'o}likowska (1985). The most
important properties of the interaction are such that a g-mode
pair gets excited if the acoustic mode amplitude exceeds certain critical
value, and that a three-mode equilibrium solution is stable
if the acoustic mode driving is weaker than g-mode damping and if the
detuning parameter is not too small. We also showed that the nonadiabacity of
the nonlinear coupling does not change the problem qualitatively. 

In the case of many g-mode pairs the equilibrium solution involving more
than two pairs in general does not exist. Instead, we have a kind
of statistical equilibrium involving many pairs exchanging energy with the
acoustic mode. We showed that such a state may be stable if
the sum of damping rates of active g-mode pairs is higher than the acoustic
mode growth rate. Moreover, these g-mode pairs should be separated in
the space of detuning parameters by more than their damping rates. The pairs
that have close detuning parameters get synchronized and interact effectively
as a single pair.

The separation of detuning parameters and the condition of the
stability imply that the range of the detuning parameters of active pairs is
of the order of the p-mode growth rate. The inverse of the growth rate
determines the timescale of the modulation of the acoustic mode amplitude.
The fast amplitude variations cause that the expression for the critical
amplitude derived in quasi-static approximation is no longer applicable and
many pairs are excited despite being stable according to the parametric
instability criterion applied to the time-averaged p-mode amplitude.

We applied our theory to a seismic model of a $\delta$~Scuti star
XX~Pyx. We consider all
unstable acoustic modes of $l\le 2$. Their dimensionless frequencies,
$\nu \sqrt{\pi/(G<\rho>)}$, are
between, $3.7$ and $6.2$, corresponding to radial orders $n>3$.
Such modes couple to g-mode pairs
that propagate either in the outer layers, $0.95\lesssim r/R\lesssim0.98$, or
in the interior, $0.1\lesssim r/R\lesssim0.5$. The two groups of g-modes
have very different properties, in particular the frequency spectra. While
the number of the inner g-modes that can be excited due to the parametric
instability is large, typically only one outer g-mode pair
has to be taken into account. The outer g-modes are orders of magnitude more
strongly damped but also much more strongly coupled to the acoustic modes,
resulting in critical amplitudes comparable to those for the inner g-modes.

The $l=2$ acoustic modes of low frequencies, $\sigma<4$, reach statistical
equilibria determined by the interactions with the inner g-modes of
$l\sim10^2$. The average amplitudes
are below the critical values for the excitation of the
outer g-modes which thus remain unexcited. The rotation causes that
the average amplitudes are by the factor of a few lower than in the case
without rotation. It is so because the rotation causes splitting of detuning
parameters and above certain rate the pairs within multiplets loose
synchronization and act independently. This increases the number of
effectively interacting g-mode pairs and leads to lower acoustic mode
amplitudes.
The effect of rotation becomes significant already at equatorial velocity
of the order of $1$ km/s. Then, the average
bolometric luminosity amplitudes are of the order of a few
milimagnitudes. The typical modulation timescales are of the order of
$10^3\div10^4$ days.

The $l=0$, $\sigma=3.909$ mode has average amplitude determined mainly by the
interaction with the outer g-mode pair, but the time behavior is affected by
the presence of a few excited inner g-mode pairs.
Effect of rotation is to increase the number of inner g-modes and these
modes take control of the acoustic mode evolution. The average amplitude is
now lower leading the outer g-modes unexcited.
The rotation again turns out to be important already at rates $\sim1$ km/s.
Then the average bolometric luminosity amplitude is about $0.07$ mag
and the modulation timescale is of the same order as in the previous case,
\ie $10^3$ days

The amplitudes of high frequency acoustic modes, $\sigma>4$, are determined
solely
by the interaction with the outer g-mode pairs. Typically, the system reaches
periodic limit cycle solutions. The rotation has no effect
in this case. The predicted bolometric luminosity amplitudes are large, much
higher than observed. The modulation timescales are of the
order of $10^2\div10^3$ days.

We found that the resonant mode coupling alone cannot be the main effect
responsible for amplitude limitation in $\delta$~Scuti stars, at least in
hotter and evolved ones as XX~Pyx.
Our calculated luminosity amplitudes are too high, in
particular for high frequency modes.
There is an obvious conflict with the observations.

A nonlinear saturation of the linear driving effect leads to a reduction
of the acoustic mode amplitudes. If strong enough it may lead to constant
amplitude pulsation such as discussed by Dziembowski and Kr{\'o}likowska
(1985). However, in our application to an evolved $\delta$~Scuti star model
the calculated amplitudes are still higher than observed. The difference
between the two results is a consequence of the difference in the internal
structure of the adopted models. We stressed that in the case when the
dominant amplitude limitation mechanism is the nonlinear saturation, the
coupling between unstable acoustic modes must be taken into account.

\Acknow{The author is very grateful to Prof. W. Dziembowski for creative and
stimulating discussions as well as for useful comments concerning the text
of this paper. This work was supported by the KBN grant No. 5 P03D 030 20.}

\end{document}